 \documentclass[12pt,a4paper,bibtotoc,liststotoc,fleqn,pointednumbers]{scrartcl}
\usepackage{latexsym}
\usepackage{graphicx}
\usepackage{cite}
\usepackage[american]{babel}
\usepackage{amsfonts}
\usepackage{amsbsy}
\title{Many-body scattering theory of  electronic systems}
\author{J.Berakdar\\
\vspace*{0.3cm}\\
Max-Planck Institut f\"ur Mikrostrukturphysik,\\
 Weinberg 2, 06120 Halle, Germany}
\begin{document}
\maketitle
\mathindent 0mm
 \noindent
\begin{abstract}
 This work reviews recent advances in the analytical treatment
of the continuum spectrum of  correlated few-body non-relativistic
 Coulomb systems.
The exactly solvable two-body problem serves as an introduction
to the non-separable three-particle system. For the latter
case we discuss the
existence of an approximate separability
of the long and the short-range dynamics which is exposed in
an appropriately chosen curvilinear coordinates.
  The three-body wave functions of the  long-ranged part of the Hamiltonian 
are derived and  methods are presented 
to account approximately for the short-ranged dynamics.
Furthermore, we present a  generalization of the methods
 employed for the  derivation of the three-body
wave functions to the scattering states of $N$ charged particles.
 To deal with thermodynamic properties of finite systems
we develop and discuss a  recent Green function me\-thodology designed
for the non-perturbative regime.
 In addition, we give a brief account on how
  thermodynamic properties and critical
phenomena can be exposed in  finite
interacting systems.
\end{abstract}
\section{{Introduction}}
The theoretical treatment of many-body systems is of 
a fundamental importance for a variety of branches in physics.
Of particular relevance to this work are highly excited
systems that consist of a finite number of
charged particles. 
Such systems are encountered in  processes involving
simultaneous excitations of few particles, e.g.   as is the case
in a  multiple ionization reaction.
From a theoretical point of view
it is fortunate that the  scattering states of   two particles  interacting
via the Coulomb potential are known exactly, for such
states  can serve 
as a  benchmark for approximate methods that are developed to deal with more
complex systems. For example, the perturbative (Born) treatment of the
two-particle Coulomb scattering and the comparison with the exact
results gives a first hint on the problems connected with  treatment of 
Coulomb scattering:
The first order term of the Born perturbation series  delivers already
the correct results for the  scattering cross section.
Higher order terms are however divergent.
This problem is mainly  traced back to the infinite range
of the Coulomb interaction that prohibits free motion in the asymptotic
regime.

In the theoretical treatment of more than two particle systems a further fundamental difficulty arises which
stems  from the inherent
non-separ\-ability of many-body interacting systems.  This
problem is of a general nature and appears basically for almost
 all forms of the
inter-\-particle interactions.

This work provides an overview on recent theoretical efforts to tackle analytically
the problems connected with the infinite-range tail of Coulomb potentials and
the non-seprabale aspects of many-body systems. As will be shown below, generally this
can be done only in an approximate way.
To deal with the long-range Coulomb interaction one dismisses the use of
 standard many-body approaches and
attempts at a direct (approximate)
 solution of the many-body non-relativistic
Schr\"odinger equation.
To expose the general feature of Coulomb potential scattering and to introduce
the basic methods used to solve the Schr\"odinger equation we start with
a brief review of the two-particle Rutherford scattering. The three-body Coulomb
scattering problem
has received recently much of attention \cite{red,bbk,jsb90,hubert,alt,prljsb,jbjsb,pra,kuni,cro,gasa,pla}.
In this work we focus only on one aspect of this problem, namely
the existence of an approximate separability
that allows to derive three-body wave functions valid in
certain region of the Hilbert space.
For the general case of a system consisting of  $N$ interacting
charged particles we derive correlated scattering states that
are to a first order in the distance exact at larger inter-particle
separation.

The wave function approach is of a limited value when it comes to the
study of 
 thermodynamic properties and critical phenomena in finite systems.
In this case detailed information on the density of states is needed.
Such information is encompassed in the many-body Green function. 
Therefore we devote a section of this work
to a general scheme for the derivation of the many-body Green operator and
briefly review a possible method to extract thermodynamic information on
finite systems starting from the Green function.

Unless otherwise stated we employ atomic units throughout and
neglect relativistic corrections.
\section{{Two charged particle scattering}}
To introduce the general frame work  and the basic notation  let us
consider
the non-relativistic scattering states of two charged particles with charges
$z_1$ and $z_2$.
The Schr\"odinger equation describing the motion in the two-particle
relative coordinate ${\bf r}$ is
\begin{equation}
\left[ -\frac{1}{2\mu}\Delta
+\frac{z_1z_2}{r} -E\right]\Psi_{\bf k}({\bf r}) =0.
\label{sch2}\end{equation}
Here ${\bf k}$ is the momentum conjugate to ${\bf r}$ and
$E=k^2/2\mu$ is the energy of the relative motion.
  $\mu$ is the reduced mass of the
two particles.
To decouple kinematics from dynamics we make the ansatz:
\begin{equation}
\Psi_{\bf k}({\bf r})=e^{i{\bf k}\cdot {\bf r}}\bar{\Psi}_{\bf k}({\bf r}).
\label{an2}\end{equation}
The distortion factor $\bar{\Psi}$ in Eq.(\ref{an2}) is solely due to the
presence of the potential.
The asymptotic properties of (\ref{sch2}) follows upon  substitution of
(\ref{an2}) in (\ref{sch2}). Then,  terms that fall off faster
than the Coulomb potential can be neglected which leads to the equation
\begin{equation}
\left[-\frac{i}{\mu}{\bf k}\cdot {\bf \nabla} +\frac{z_1z_2}{r}
\right] \bar{\Psi}_{\bf k}({\bf r})=0.
\label{as2}\end{equation}
This equation can be solved by the ansatz $\bar{\Psi}=\exp(i\phi)$.
Upon insertion in Eq.(\ref{as2}) this ansatz yields
\begin{equation}
\phi^\pm_{\bf k}({\bf r})=\pm\frac{z_1z_2\mu}{k}\ln a(r\mp
\hat{\bf k}\cdot {\bf r}).
\label{phase2}\end{equation}
The factor $z_1z_2\mu/k$ is called the Sommerfeld parameter and
characterizes 
the strength of the interaction. The integration constant
$a$ has a dimension of a reciprocal length and the value
$a=k$. The important point here is that  the natural  coordinate
that appears in the treatment of  Coulomb scattering is the so-called
{\em parabolic} coordinate $\xi^\pm:=r\mp \hat{\bf k}\cdot {\bf r}$
where the $+$ or $-$
corresponds respectively to
incoming or outgoing-wave boundary conditions.
\section{{The three-particle coulomb continuum states}}
%
%
%
%
In contrast to the two-body problem, an exact
 derivation of the three-body quantum states is not possible.
Nonetheless, under certain (asymptotic) assumptions
analytical solutions can be obtained that contain some general features 
of the two-body scattering,
such as the characteristic asymptotic phases.
As in the preceding section the center-of-mass motion
of a three-body system can be factored out. The internal motion 
of the three charged particles with masses $m_i$ and 
charges $z_i\ ; i\in {1,2,3}$ can be  described by  one set
 of the three
Jacobi coordinates $({\bf r}_{ij},{\bf R}_k);\ i,j,k\in\{ 1,2,3\} 
;\epsilon_{ijk}\neq 0; \ j>i$. Here ${\bf r}_{ij}$ is the relative
internal separation of the pair $ij$ and ${\bf R}_k$ is the
position of the third particle ($k$) with respect to the center of mass of the
pair $ij$.
The three sets of Jacobi coordinates are connected with each other via the transformation
\begin{eqnarray}
\left( \begin{array}{c}
{{\bf r}_3} \\ {\bf R}_2 \end{array}\right) ={\mbox{{\bf D}}_1}
\left( \begin{array}{c}{{\bf r}_{23}} \\ {{\bf R}_1} \end{array}\right)
\hspace*{0.3cm}\mbox{and}\hspace*{0.3cm}
\left( \begin{array}{c}
{{\bf r}_{12}} \\ {{\bf R}_3} \end{array}\right) ={\mbox{{\bf D}}_2}
\left( \begin{array}{c}{{\bf r}_{23}} \\ {{\bf R}_1} \end{array}\right)
 \label{jacob}\ \end{eqnarray}
where 
\begin{eqnarray}
{\mbox{{\bf D}}_1} &=& \left( \begin{array}{cc} \mu_{23}/m_3 & 1\\ 1-
\mu_{13}\cdot\mu_{23}/m_3^2 & -\mu_{13}/m_3 \end{array}\right)\nonumber\\
\nonumber\\
{\mbox{{\bf D}}_2} &=& \left( \begin{array}{cc} -\mu_{23}/m_2 & 1\\ -1+
\mu_{12}\cdot\mu_{23}/m_2^2 & -\mu_{12}/m_2 \end{array}\right)
\ \end{eqnarray}
The reduced masses are defined as  $\mu_{ij}=m_im_j/(m_i+m_j)\ ;
\ i,j\in \{ 1,2,3 \};\ j>i$. Accordingly, the  momenta  conjugate  to
(${\bf r}_{ij},{\bf R}_k$) are defined as $({\bf k}_{ij},{\bf K}_k)$. These
momenta are related to   each other by
\begin{eqnarray}
\left( \begin{array}{c}
{{\bf k}_{23}}\\ {\bf K}_1\end{array}\right) &=&{\mbox{{\bf D}}_2}^t
\left( \begin{array}{c}{{\bf k}_{12}} \\ {{\bf K}_3}\end{array}\right)
=
{\mbox{{\bf D}}_1}^t
\left( \begin{array}{c}{{\bf k}_{13}} \\ {\bf K}_2
\end{array}\right),\ \end{eqnarray}
where ${\mbox{{\bf D}}_1}^t$ and ${\mbox{{\bf D}}_2}^t$ are 
 transposed matrices of 
 ${\mbox{{\bf D}}_1}$ and ${\mbox{{\bf D}}_2}$,
 respectively. The scalar product    
$ ({\bf r}_{ij},{\bf R}_k)\cdot \left( \begin{array}{c}
{\bf k}_{ij}\\ {\bf K}_k \end{array}\right) $ is invariant
 for all three sets of 
Jacobi coordinates. The kinetic energy operator $H_0$ 
is then diagonal and reads
\begin{eqnarray}
H_0=-\frac{1}{2\mu_{ij}}{\mit\Delta}_{{\bf r}_{ij}}
-\frac{1}{2\mu_{k}}{\mit\Delta}_{{\bf R}_k}\hspace*{1.0cm}
\forall\hspace*{0.3cm} ({\bf r}_{ij},{\bf R}_k)
\ ,\label{frei}\ \end{eqnarray}
where   $\mu_k=m_k(m_i+m_j)/(m_1+m_2+m_3)$.
The eigenenergy of (\ref{frei}) is then given as
\begin{eqnarray}
E_0=\frac{{\bf k}_{ij}^2}{2\mu_{ij}}
+\frac{{\bf K}_k^2}{2\mu_{k}}
\hspace*{1.0cm}\forall\hspace*{0.3cm} ({\bf r}_{ij},{\bf R}_k)\ .
\label{e0}\ \end{eqnarray}
The time-independent Schr\"odinger equation of the system reads
\begin{eqnarray}
\left[ H_0 + \sum^3_{\stackrel{i,j}{j>i}}\frac{z_{ij}}{r_{ij}}-E\right]
 \ \langle {\bf r}_{kl},{\bf R}_{m}|\Psi_{{\bf k}_{kl},{\bf K}_{m}}  \rangle =0\ .
\label{schraa}\ \end{eqnarray}
Here we defined the product charges $z_{ij}=z_iz_j$;  $ j>i\in\{ 1,2,3\}$. 
The relative coordinates  
 $r_{ij}$ occurring in the  Coulomb potentials have to be
expressed in terms of the appropriately chosen Jacobi-coordinate set
 (${\bf r}_{kl},{\bf R}_{m}$).\\
 Asymptotic  scattering solutions of (\ref{schraa}) for large
interparticle distances ${\bf r}_{ij}$
have the form \cite{red,peter2,merk,bbk,hubert}: 
\begin{eqnarray}
\smash{\lim_{\stackrel{r_{ij}\to\infty}{R_{k}\to\infty}}}
\Psi_{{\bf k}_{ij},{\bf K}_k}({\bf r}_{ij},{\bf R}_k) &\to&
   (2 \pi )^{- 3} \exp(i\  {{\bf k}_{ij}} \cdot {{\bf r}_{ij}}
    + i\  {{\bf K}_k} \cdot {{\bf R}_k}  )
 \nonumber\\
&&\times \ \prod_{\stackrel{i,j=1}{j>i}}^3\exp
\left(\pm i{\alpha}_{ij}\ln(k_{ij}\ r_{ij}\pm
{\bf k}_{ij}\cdot{\bf r}_{ij})\frac{ }{}\right)\ ,
\hspace*{0.3cm}\forall\hspace*{0.2cm}({\bf r}_{ij},R_k).\nonumber\\
\label{red}\ \end{eqnarray}
The '$+$'  and '$-$' signs refer to 
 outgoing and incoming
boundary conditions, respectively.
Similarly to the two-body problem, the Sommerfeld-parameter ${\alpha}_{ij}$ are given by
\begin{eqnarray}
{\alpha}_{ij}=\frac{Z_{ij}\mu_{ij}}{k_{ij}}
.\label{somp2}\ \end{eqnarray}
The asymptotic state (\ref{red}) is a straightforward generalization 
of the two-body asymptotic given by Eq.(\ref{an2})  
 to three-body systems. However, unlike
the situation in two-body scattering, in  three-body systems other
types of asymptotic are present where, in a certain set
(${\bf r}_{ij},{\bf R}_k$), one Jacobi coordinate
tends to infinity whereas the other coordinate remains finite
\cite{alt}. 
%
%
The asymptotic states (\ref{red}) serve as the boundary conditions 
that have to be satisfied
by the scattering solutions of the Schr\"odinger equation. The derivation of these
solutions is a delicate task and will be the subject of the
remainder of this section.

The general approach here is to consider the three-body system as the subsume of three
non-interacting two-body subsystems \cite{pra}. Since we know the appropriate
coordinates for each of these two-body subsystems (the parabolic coordinates)
we formulate the three-body problem in a similar coordinate frame with
\begin{eqnarray} &&\{ \xi_k^\mp = r_{ij} \pm \ {\bf \hat
 {k}}_{ij}\cdot{\bf r}_{ij} \},\:
  \epsilon_{ijk}\ne 0;\ j>i, k\in
 [1,3], \label{cor}\end{eqnarray}
 where ${\bf \hat{k}}_{ij}$ denote
 the directions of the
 momenta ${\bf k}_{ij}$.
 Since we are dealing with a six-dimensional problem three other
 independent coordinates are needed in addition to (\ref{cor}).
 To make a reasonable choice for these remaining coordinates we remark
 that, usually, the momenta ${\bf k}_{ij}$ are
 determined experimentally, i. e. they can be considered as the
 laboratory-fixed coordinates. In fact it can be shown
 that the coordinates (\ref{cor}) are related to the Euler angles.
 Thus, it is advantageous to choose body-fixed coordinates.
 Those are conveniently chosen as
 \begin{eqnarray}
\{ \xi_k = r_{ij} \},
&& \epsilon_{ijk}\ne 0;\ j>i, k\in
 [4,6]. \label{cor1}\end{eqnarray}
 Upon a mathematical analysis it can be shown that the
 coordinates (\ref{cor},\ref{cor1}) are linearly independent \cite{pra}
 except for some singular points where the Jacobi determinant
 vanishes.
 The main task is now to rewrite the three-body Hamiltonian in the
 coordinates (\ref{cor},\ref{cor1}). 
To this end it is useful to
 factor out the
 trivial plane-wave part [as done in Eq.(\ref{an2})] by making the 
ansatz
\begin{eqnarray}
\Psi_{{\bf k}_{ij},{\bf K}_k} ( {\bf r}_{ij},{\bf R}_k )  = N\,
 \exp(i\ {\bf r}_{ij}\cdot{\bf k}_{ij} + i\ {\bf R}_{k}\cdot{\bf K}_{k})\ 
\overline{\Psi}_{{\bf k}_{ij},{\bf K}_k} ({\bf r}_{ij},{\bf R}_k).
\label{ans}\ \end{eqnarray}
Inserting the
ansatz (\ref{ans}) into the Schr\"odinger Eq.(\ref{schraa}) 
leads to the equation
\begin{eqnarray}
\left[ \frac{1}{\mu_{ij}}{\mit\Delta}_{{{\bf r}_{}}_{ij}}
+\frac{1}{\mu_k}{\mit\Delta}_{{{\bf R}_{}}_k}+
 2i\left( \frac{1}{\mu_{ij}}{\bf k}_{ij}\cdot
{\bf \nabla}_{{{\bf r}_{}}_{ij}}\ + \frac{1}{\mu_k} {\bf K}_{k}\cdot
{\bf \nabla}_{{{\bf R}_{}}_k}\right) - 2 \sum^3_{\stackrel{m,n}{n>m}}
\frac{Z_{ij}}{{r}_{mn}}\right]  \overline{\Psi}({{\bf r}_{}}_{ij}
,{{\bf R}_{}}_k)
=0.\nonumber\hspace*{-1cm}\\
\label{mschra}\ \end{eqnarray}
In terms of the coordinates (\ref{cor},\ref{cor1}) Eq.(\ref{mschra}) casts
\begin{eqnarray}
H  \overline{\Psi}_{{\bf k}_{ij},{\bf K}_k} (\xi_1,\dots,\xi_6)
= \left[\ H_{\rm par}+H_{\rm in}+H_{\rm mix}\right]
\overline{\Psi}_{{\bf k}_{ij},{\bf K}_k} (\xi_1,\dots,\xi_6)=0
\ .\label{h}\ \end{eqnarray}
%
%
The operator $H_{\it par}$ is differential in the
 {\em para\-bolic} coordinates $\xi_{1,2,3}$ only whereas
 $H_{\it int}$ acts on the
 internal degrees of freedom $\xi_{4,5,6}$.
 The mixing term $H_{\it mix}$ arises from the
  off-diagonal elements of the metric tensor
  and plays the role of a rotational coupling in a hyperspherical
  treatment.\\
  The essential point is that the differential operators
  $H_{\it par }$ and $H_{\it int}$
   are exactly separable in the coordinates $\xi_{1\cdots 3}$
   and $\xi_{4\cdots 6}$, respectively, for  they can be written as \cite{pra}
\begin{eqnarray}
H_{\it par}=\sum^3_{j=1}H_{\xi_j}\, ;\, \:
[H_{\xi_j},H_{\xi_i}]=0;\:  \forall\  i,j\in\{1,2,3\},
\label{comp}\end{eqnarray}
and
\begin{eqnarray}
H_{\it int}=\sum^6_{j=4}H_{\xi_j}\, ;\, \:
[H_{\xi_j},H_{\xi_i}]=0 ;\: \forall\  i,j\in\{4,5,6\},
\label{comi}\end{eqnarray}
where
\begin{eqnarray}
H_{\xi_j}&=&\frac{2}{\mu_{lm}r_{lm}}\left[{\partial}_{\xi_j}\ \xi_j\
 {\partial}_{\xi_j}
 +i k_{lm}\ \xi_j\ {\partial}_{\xi_j}
 - \mu_{lm}\ z_{lm}\right]; \nonumber\\
&&\epsilon_{jlm}\neq 0, \ j\in\{1,2,3\}.
 \label{Hj}\end{eqnarray}
 and
 \begin{eqnarray}
 H_{\xi_4}&=&\frac{1}{\mu_{23}}\left[\frac{1}{\xi^2_4}{\partial}_{\xi_4}\
 \xi_4^2\ {\partial}_{\xi_4} +i2 k_{23}\ \frac{\xi_1-\xi_4}{\xi_4}
   {\partial}_{\xi_4}\right]; \nonumber\\ \label{xi_4}\\
 H_{\xi_5}&=&\frac{1}{\mu_{13}}\left[\frac{1}{\xi^2_5}{\partial}_{\xi_5}\
 \xi_5^2\ {\partial}_{\xi_5} +i2 k_{13}\ \frac{\xi_2-\xi_5}{\xi_5}
   {\partial}_{\xi_5}\right]; \nonumber\\ \label{xi_5}\\
 H_{\xi_6}&=&\frac{1}{\mu_{12}}\left[\frac{1}{\xi^2_6}{\partial}_{\xi_6}\
 \xi_6^2\ {\partial}_{\xi_6} +i2 k_{12}\ \frac{\xi_3-\xi_6}{\xi_6}
   {\partial}_{\xi_6}\right].\nonumber\\
     \label{xi_6}\end{eqnarray}
  The operator $H_{\it mix}=H-H_{\it par}-H_{\it int}$ derives
  from the expression
  \begin{eqnarray}
  H_{\it mix}:=&
   \sum_{u\neq v=1}^6&\left\{
   ({\bf \nabla}_{{{\bf r}_{}}_{ij}}\xi_u)\cdot
   ({\bf \nabla}_{{{\bf r}_{}}_{ij}}\xi_v )
   +({\bf \nabla}_{{{\bf R}_{}}_{k}}\xi_u)\cdot
   ({\bf \nabla}_{{{\bf R}_{}}_{k}}\xi_v)\right\}{\partial}_{\xi_u}
   {\partial}_{\xi_v}.\nonumber\\
   \label{lmix}\ \end{eqnarray}
  Noting that $H_{\xi_j}, j=1,2,3$ is simply the
  Schr\"odinger operator
  for the two-body scattering  rewritten in parabolic coordinates
  (after factoring out the plane-wave part),
  one arrives immediately, as a
  consequence of Eq.(\ref{comp}),
   at an expression for the
  three-body wave function as a product of three two-body continuum waves
with the correct boundary conditions (\ref{red}) at large inter-particle separations.
This result is valid
  if the contributions of  $H_{\it int}$ and $H_{\it mix}$ are negligible
as compared to $H_{\it par}$, which is in fact the case
for large interparticle separations \cite{pra} {\em or} at high
particles' energies.

It should be noted that this same result can 
deduced  in a Jacobi coordinate system however the operators
 $H_{\it par}$, $H_{\it int}$ and $H_{\it mix}$ have a much more complex
 representation in the
 Jacobi coordinates (cf. Ref.\cite{pra}).

With decomposing the total Hamiltonian in $H_{\it par}$, $H_{\it
int}$ and $H_{\it mix}$ we achieved a result similar to that
obtained by Pines \cite{pines} and co-worker for  the interacting electron gas: 
The system
is decomposed into a long-range and a short range components
described respectively by $H_{\it par}$ and $H_{\it int}$. As
clear from Eq.(\ref{red}), the eigenfunctions of $H_{\it par}$ have an oscillatory
asymptotic behaviour whereas the eigenstates of $H_{\it int}$ decay
for large interparticle distances \cite{unpub}.  The mixing
term $H_{\it mix}$ couples the short-range to the long-range modes
of the system.

The analytical structure of the Eqs.(\ref{comp}-\ref{lmix}) deserves
several remarks:
\begin{itemize}
\item The total potential is contained in the operator $H_{\it par}$,
as can be seen from Eqs.(\ref{Hj}). Thus, the eigenstates of $H_{\it par}$
treats  the total potential in an  exact manner. This means on the other hand
that the operators $H_{\it int}$ and $H_{\it mix}$ are parts of the kinetic
energy operator. This situation is to be contrasted with other treatments
 \cite{wan,peter1,peter2,rau71,klar,read,jim,rost,macek}
 of the three-body problem in regions of the space space where the
potential is smooth, e.g. near a saddle point. In this case one usually
expands the potential around the fix point and accounts for the kinetic energy
in an exact manner.
\item
In Eq.(\ref{Hj}) the total potential appears as a sum of three
two-body potentials. It should be stressed, that this splitting is
arbitrary, since the dynamics is controlled by the total
potential. I.e., any other splitting that leaves the total
potential invariant is equally justified. This fact we will use
below for the construction of three-body states. For large
inter-particle separation the operators $H_{\it int}$ and $H_{\it
mix}$ are negligible as compared to $H_{\it par}$ and the
splitting of the total potential as done in Eqs.(\ref{Hj}) becomes
uni\-que. This means, for large particles' separation the three-body
dynamics is controlled by sequential  two-body scattering events.
\item The momentum vectors ${\bf k}_{ij}$ enter the Schr\"o\-dinger
equation via the
asymptotic boundary conditions. Thus, their physical meaning, as
two-body relative momenta,  is restricted to the asymptotic region
of large inter-particle distances. The consequence of this
conclusion is that, in general, any combination or functional form
of the momenta ${\bf k}_{ij}$ is legitimate as long as the total
energy is conserved  and the boundary conditions are fulfilled
(the energies and the wave vectors are liked via a parabolic
dispersion relation).
 This fact has been employed
in Ref.\cite{alt}  to constructed three-body wave functions
with position-dependent momenta ${\bf k}_{ij}$ and in  Ref.\cite{exch}
to account for off-shell transitions.
\item
The separability of the operators (\ref{xi_4}-\ref{xi_6}) may be used to
deduce representations
of  three-body states  \cite{unpub} that
diagonalize simultaneously $H_{\it par}$ and $H_{\it int}$.
 It should be noted however,
that generally
 the operator $H_{\it mix}$, which has to be neglected in this case,
 falls off with distance as fast as $H_{\it int}$.
\item
As well-known, each separability of a system implies   a related
conserved quantity. In the present case we can only speak of an
approximate separability and hence
of approximate conserved quantum numbers.\\
If we discard $H_{\it int}$ and $H_{\it mix}$ in favor of $H_{\it
par}$, which is justified for $k_{ij}\xi_k,\; \epsilon_{ijk}\neq 0,
k\in [1,3]$ (i.e.~for a large $\xi_k$ or for a high two particle
momentum $k_{ij}$), the three-body good quantum numbers are
related to those in a two-body system in parabolic coordinates.
The latter are the two-body energy, the eigenvalue of the
  component of the Lenz-Runge operator along a quantization axis $z$ and
the eigenvalue of the component of the angular momentum operator along $z$.
 In our case the quantization axis $z$
 is given by the linear momentum direction $\hat {\bf k}_{ij}$.\\

In Ref.\cite{pra} the three-body problem has been formulated
in hyperspherical-para\-bolic coordinates.
In this case the operator $H_{\it int}$ takes on the form of the
grand angular momentum operator. This observation is useful to expose the
relevant angular momentum quantum numbers in case $H_{\it mix}$ can be neglected.
\item
In Ref.\cite{hubert,dipl} the three-body system has been expressed
in the coordinates $\eta_j=\xi_j^+,\; j=1,2,3$ and $\bar
\eta_j=\xi_j^-,\; j=1,2,3$.  This is the direct extension of the
para\-bolic coordinates for the body problem (cf. Section 3)
 to the three-body
problem. From a physical point of view this choice is not quite
suited, for scattering states are sufficiently quantified by
outgoing or incoming wave boundary conditions (in contrast to
standing waves, such as bound states whose representation
requires a combination
of incoming and outgoing waves). Therefore, to account for
the boundary conditions in scattering problems, either the
coordinates $\eta_j$ or $\bar \eta_j$ are needed.
The appropriate
choice of the remaining three coordinates should be made on the
basis of   the form of the forces governing  the three-body
system. In the present case where external fields are absent we
have chosen $\xi_k=r_{ij}, \; k=4,5,6$ as the natural coordinates
adopted to  the potential energy operator.
\end{itemize}
\subsection{{Coupling the short and the long-range dynamics}}
In the preceding sections we pointed out that
the eigenstates of $H_{\it par}$ can be deduced analytically.
 These eigenfunctions,
  even though are well defined in the entire Hilbert space,
 constitute a justifiable approximation to the exact three-body
state in the asymptotic region only
(e.g.~in the region of larger inter-particle
separation or for higher energies).
 This fact is important when
it comes
   to  evaluating reaction amplitudes, for such amplitudes
involve the many-body scattering state in the entire Hilbert
space. Therefore, an adequate description of the short-range
dynamics may be necessary, in particular in cases where the
contributions to the matrix elements of the transition amplitudes
originates from the internal region, i.e. when the reaction takes
place at small interparticle distances. Nevertheless the
eigenstates of $H_{\it par}$ can, and have been  used for the
calculations of transition matrix elements, as for example done
below. In this case the justification of this doing must go beyond
the asymptotic correctness argument.

In this section we seek  three-body wave functions
   that diagonalize,
in addition to $H_{\it par}$,
  parts of $H_{\it int}$ and $H_{\it mix}$.

One method that turned out to be particularly
effective for this purpose  relies on the observations: a) In a three-body system
the form of the two-body potentials $z_{ij}/r_{ij}$ are generally
irrelevant, as long as the total potential is conserved. c) To
keep the mathematical structure of the operators
(\ref{comp},\ref{Hj}) unchanged and to introduce a splitting of
the total potential while maintaining
 the total potential's rotational
invariance one can assume the strength of the individual
two-body interactions, characterized by $z_{ij}$, to be
dependent on $\xi_{4,5,6}$.
This means we introduce position dependent product charges as
\begin{equation}
\bar{z}_{ij}=\bar{z}_{ij}(\xi_4,\xi_5,\xi_6),
\label{z}\end{equation}
with
\begin{equation}
\sum_{j>i=1}^3\frac{\bar{z}_{ij}}{r_{ij}}=\sum_{j>i=1}^3
\frac{{z}_{ij}}{r_{ij}}
.\end{equation}
To obtain  the {\em many-body} potentials
$\bar{V}_{ij}:={\bar{z}_{ij}}/{r_{ij}}$
 we express them as
a linear mixing of  the isolated {\em two-body} interactions
$V_{ij}:={{z}_{ij}}/{r_{ij}}$, i.~e.
\begin{eqnarray}
\left( \begin{array}{ccc}
\overline{V}_{23} \\ \overline{V}_{13} \\ \overline{V}_{12}
 \end{array}\right) ={{\cal A}}
 \left( \begin{array}{ccc}
 V_{23} \\ V_{13} \\ V_{12} \end{array}\right) \ ,
 \label{trans}\ \end{eqnarray}
 where ${\cal A}(\xi_4,\xi_5,\xi_6)$ is a $3\times 3$ matrix.
The matrix elements are then  determined according to
1) the properties of the total potential surface, 2) to reproduce
the correct asymptotic of the  three-body states and 3) in a way that
minimizes $H_{\it int}$ and $H_{\it mix}$. It should be stressed
that the procedure until this stage  is exact. It is merely
a splitting of the total potential that leaves
this potential and hence
the three-body
Schr\"odinger equation unchanged.

\subsection{An electron pair in the field of a positive ion}
To be specific let us  demonstrate the method  for the case of two
electrons moving in the Coulomb field of a residual ions. This
brings about some simplifications since the ion can be considered
infinitely heavy as compared to the electron mass. Traditionally,
the electrons are labeled $a$ and $b$ and their positions and
momenta with respect to the residual ion are respectively called
 ${\bf r}_a$, ${\bf r}_b$ and  ${\bf k}_a$, ${\bf k}_b$.
Adopting this notation, the eigenstate of the operator $H_{\it par}$
reads:
\begin{eqnarray}
 \overline{\Psi}_{{\mathbf k}_{a},{\mathbf k}_{b}}
 (\xi_{1\cdots 6})&=&
\mbox{ }
       _1F_1 \left( i\beta_{a}, 1,\ -i k_{a}\ \xi_1
        \ \right)
\mbox{ }\nonumber\\
&& \mbox{ }
       _1F_1 \left( i\beta_{b},\ 1,\ -i k_{b}\ \xi_2
        \ \right)\nonumber\\
\mbox{ }
 &&      _1F_1 \left( i\beta_{ab},\ 1,\ -i k_{ab}\ \xi_3
  \ \right).\nonumber\\
\label{ds3c}\  \end{eqnarray}
 Here we denoted  the  relative electron-electron
 momentum by ${\bf k}_{ab}=\frac{1}{2} ({\bf k}_{a}-
{\bf k}_{b})$ whereas
$_1F_1[a,b,x]$ stands for the confluent hypergeometric function and
$\beta_{j},\ j\in\{a,b,ab\}$ are
the Sommerfeld parameters
\begin{equation}
\beta_{j}=\frac{\bar z_{j}}{v_{j}},\ j\in
\{a,b,ab\} ,\label{som}\end{equation}
with $v_j$ being  the velocities corresponding to the momenta ${\mathbf k}_{j}$ and
$\bar z_j,\ j\in\{a,b,ab\}$ are the electrons-ion and electron-electron
effective product charges, respectively. The form of $\bar z_j$ is still to be determined.

 Below 
the functions $\bar z_{j}(\xi_{4\cdots 6})$ are given
 that preserve the total potential, possess
 the correct three-body asymptotic
and incorporate features of the many-particle motion at the complete
fragmentation threshold, namely $ $ along the saddle point of the
total potential, the so-called Wannier ridge
 \cite{wan,peter1,peter2,rau71,klar,read,jim,rost,macek}.
Since $\bar z_j$ are assumed to depend on the internal coordinates only
($\xi_4,\xi_5,\xi_6$) the wave function (\ref{ds3c}) is still an
eigenstate of the long-range Hamiltonian $H_{\it par}$ (given by Eq.\ref{Hj}). In physical
terms it can be said that the effect of the short-range part of the
Hamiltonian is to modify dynamically the  coupling
strength of the isolated two particle system ($z_{ij}$).

 To ensure the invariance of the
 Schr\"odinger equation  under the introduction of the
product charges $\bar z_{j}(\xi_{4\cdots 6})$ the three conditions
\begin{eqnarray}
\sum_{j}\frac{\bar z_{j}(\xi_{4\cdots 6})}{r_j}
&\equiv&\frac{-z}{r_a}+\frac{-z}{r_b}+\frac{1}{r_{ab}},\:
j\in\{a,b,ab\}, \label{potin}\end{eqnarray}
 have to be fulfilled
(${\bf r}_{ab}$ is the electron-electron relative coordinate and
$z$ is the charge of the residual ion). The wave functions
containing $\bar z_j$  must be compatible with the three-body
asymptotic boundary conditions. These are specified by the shape
and size of the triangle formed by the three particles (two
electrons and the ion): I. e., the derived wave function must be,
to a leading order, an asymptotic solution of the three-body
Schr\"odinger equation when the aforementioned triangle tends to a
line (two particles are close to each other and far away from the
third particle) or in the case where, for an arbitrary shape, the
size of this triangle becomes infinite. The latter limit implies
that all interparticle coordinates $r_{a,b,ab}$ must grow with the
same order, otherwise we eventually fall back to the limit of the
three-particle triangle being reduced to a line \cite{pra}, as
described above. In addition we require the Wannier threshold law
for double electron escape to be reproduced when the derived wave functions
are
used for the evaluation of the matrix elements.\\
All of the above conditions are sufficient to
determine $\bar z_j$ and thus the wave function (\ref{ds3c}). This wave
function is called "dynamically screened three-body Coulomb  wave
function"  $\Psi_{\rm DS3C}$. This is because this wave function
consists formally of three Coulomb waves $ $ where the short-range
dynamics enters as a dynamical screening of the strength of the
two-body interaction.

 The applicability of the  wave function
$\Psi_{\rm DS3C}$  to scattering reactions is hampered by the
involved functional dependence  leading to complications in
the numerical determination of the normalization and of the
scattering matrix elements.
 Furthermore, the
incorporation of the three-body scattering dynamics at shorter
distances brings about  intrinsic practical
disadvantages as compared  to an  approach where
 $\bar z_j$ are constant (this approach is usually called three-body Coulomb wave
 method , 3C). Namely,
the construction of $\Psi_{\rm DS3C}$ has to be individually
undertaken for
 given charge and mass states of the specific three particle system at hand.
This is comprehensible since properties of the
total potential  are inherent to the particular three-body system under
investigation.

 As shown in Ref. \cite{bbk91} the normalization of the
3C wave function is readily determined from the asymptotic flux.
 This procedure
has not been accessible in the case of $\Psi_{\rm DS3C}$ due to
the position dependence of $\bar z_j$.

 To overcome this difficulty (and that associated with the
six-dimensional numerical integration when evaluating transition
matrix elements) we note that the position dependence of
$\bar z_j(r_a,r_b,r_{ab})$  occurs (due to dimensionality
considerations) through ratios of the interparticle distances.
Thus, this dependence can be converted into velocity dependence by
assuming that
\begin{equation} \frac{r_i}{r_j}\propto \frac{v_i}{v_j}.
\label{prop}\end{equation}
The proportionality constant in Eq. (\ref{prop}) could be of an arbitrary
functional dependence. It should be emphasized  that the  approximation
(\ref{prop}) is not a classical one, i.~e. it is not assumed that
the particles' motions proceed along classical trajectories [conversely, if the
motion were classically free, Eq. (\ref{prop}) holds].
It merely means that the total potential is exactly
diagonalized in the phase space where Eq. (\ref{prop}) is
satisfied, as readily deduced from Eq. (\ref{potin}).\\
 Eq. (\ref{prop}) renders possible the normalization
of $\Psi_{\rm DS3C}$
 since in this case we obtain $\bar z_j=\bar z_j(k_a,k_b,k_{ab})$ and the arguments
used in Ref. \cite{bbk91} can be repeated to deduce for
the normalization $N$ the expression
\begin{eqnarray}
N&=&\prod_jN_j,\: j\in \{a,b,ba\}\nonumber\\
 N_j&=&\exp[-\beta_{j}(k_a,k_b,k_{ba})\pi/2] \,
\Gamma[1-i\beta_{j}(k_a,k_b,k_{ba})].\\ \nonumber
\label{normend}\end{eqnarray}
Here $\Gamma(x)$ is the Gamma function. The
velocity-dependent product charges \cite{aust} have the form
\begin{eqnarray}
\bar z_{ba}({\mathbf v}_a,{\mathbf v}_b )&=&
 \left[ 1-(f\ g)^2\  a^{b_1}\right]a^{b_2}\label{zab}\\
\bar z_{a}({\mathbf v}_a,{\mathbf v}_b )&=&
 -1 + (1 -\bar z_{ba})\frac{v_a^{1+a}}{(v_a^a+v_b^a)v_{ab}} \label{za}\nonumber\\
 \\
\bar z_{b}({\mathbf v}_a,{\mathbf v}_b )&=&
 -1 + (1 -\bar z_{ba})\frac{v_b^{1+a}}{(v_a^a+v_b^a)v_{ab}}. \nonumber\\
 \label{zb}
 \end{eqnarray}
The functions occurring in  Eqs.\ (\ref{zab},\ref{za}) are defined as
(${\bf v}_a,\, {\bf v}_b$ are the electrons' velocities and ${\bf v}_{ab}=
{\bf v}_a - {\bf v}_b$)
\begin{eqnarray}
f&:=& \frac{3+\cos^24\alpha}{4},\hspace*{0.5cm} \tan\alpha=\frac{v_a}{v_b}\\
g&:=& \frac{v_{ab}}{v_a+v_b}\\
b_1&:=& \frac{2v_av_b\cos(\theta_{ab}/2)}{v_a^2+v_b^2}\\
b_2&:=& g^2(-0.5 + \bar\mu)\\
a&:=& \frac{E}{E+0.5}, \label{aa}
\end{eqnarray}
where  $E$  is being measured in atomic units and $\bar\mu$ is the
Wannier index (the value of $\bar\mu$ depends on the residual ion
charge value, the numerical value of
$\bar\mu$ for a unity charge of the residual ion is $\bar\mu=1.127$). The
interelectronic relative angle $\theta_{ab}$ is given by
  $\theta_{ab}:=\cos^{-1} \hat{\mathbf  v}_{a} \cdot  \hat {\mathbf v}_{b}$.
In case of higher excess energies ($E\gg 1$) it is readily verified that
$a\to 1$ [Eq. (\ref{aa})] and all modifications of the charges (\ref{zab}-\ref{zb})
which are due to incorporating the  Wannier threshold law become irrelevant.
The charges (\ref{zab}-\ref{zb}) reduce then to those given in Ref. \cite{pra}
with Eq. (\ref{prop}) being applied.
From the functional forms of the charges
(\ref{zab}-\ref{zb}) it is clear that when
two particles approach each other (in velocity space) they experience their full
two-body Coulomb interactions, whereas the third
one `sees' a net charge equal
to the sum of the charges of the two close particles.
\subsection{Applications to atomic scattering problems}
In this section we assess the analytical methods developed above
by performing a numerical evaluation of many-body scattering amplitudes.
The reaction we are considering here is the electron-impact ionization of
atomic hydrogen. In the final channel of this collision process two interacting
electrons move in the double continuum of a residual ion. Hence a correlated three-body
wave function is needed to represent this state.
For this wave function we employ the approximate expressions given in the
preceding sections.  The initial state consists of an incoming single-particle
wave that represents the projectile electron and a bound state of atomic hydrogen.

The complete information on this
reaction is obtained by measuring the coincidence rate for the
emission of two continuum electrons with specified wave vectors, i.e.
the energies $E_a,E_b$  and the emission solid angles $\Omega_a,\Omega_b$
 of the two electrons are determined for a given
incident energy of the projectile electron. Due to energy conservation
it suffices to determine the energy of one of the electrons. Therefore,
one measures in this way a
triply differential cross section (TDCS), i.e. a  cross
section differential $\Omega_a,\, \Omega_b$ and $E_b$. 

If the spin of the electrons is not resolved, the TDCS
is a statistically weighted average of singlet and triplet scattering cross sections
\begin{eqnarray}
 {\rm TDCS}( {\mathbf k}_a ,{\mathbf k}_b )= c   \left(
\frac{1}{4}|T^s|^2  + \frac{3}{4}|T^t|^2\right)
\label{tdcs}\end{eqnarray}
where ${\mathbf k}_{i}$ is the momentum of the incident projectile
and $c=(2\pi)^4 (k_a\, k_b)/{k_i} $.
The singlet $T^s$ and triplet transition matrix elements   $T^t$
derive  from the corresponding  transition operators ${\cal T}^s$
 and ${\cal T}^t$, where
\begin{eqnarray}
{\cal T}^s&=&({\rm I} + {\cal P}_{ab}){\cal
 T}_{fi}({\mathbf k}_a ,{\mathbf k}_b )\nonumber\\
{\cal T}^t&=&({\rm I} - {\cal P}_{ab}){\cal
 T}_{fi}({\mathbf k}_a ,{\mathbf k}_b ).
\label{sintri}\end{eqnarray}
The action of the  exchange operator ${\cal P}_{ab}$
on the operator ${\cal T}_{fi}$   is given by
${\cal P}_{ab}{\cal T}_{fi}({\mathbf
 k}_a ,{\mathbf k}_b )={\cal T}_{fi}({\mathbf k}_b ,{\mathbf k}_a )$.
The prior
representation of
${\cal T}_{fi}({\mathbf k}_a ,{\mathbf k}_b )$   is given by
\begin{eqnarray}
 T_{fi}({\mathbf k}_a ,{\mathbf k}_b )=\langle\Psi|V_i|\Phi_{{\mathbf k}_{i}}\rangle
\ .\label{t}\end{eqnarray}
The wave function  $\Psi$ is obtained from   Eq. (\ref{ds3c}) as
\[
\Psi_{{\bf k}_a,{\bf k}_a}=
N\exp{i({\bf k}_a\cdot{\bf r}_a +{\bf k}_b\cdot{\bf r}_b)}\bar \Psi_{{\bf k}_a,{\bf k}_a}.\]
The  three-body system in the initial channel
is described by $|\Phi_{{\mathbf k}_{i}}\rangle$.
Assuming $|\Phi_{{\mathbf k}_{i}}\rangle$
to be the asymptotic initial-state, i. e.
$\langle {\mathbf r}_a,{\mathbf r}_b|\Phi_{{\mathbf k}_{i}}\rangle$
is a product of an
incoming plane wave representing the incident projectile  electron and
 an undistorted $1s$-state of atomic hydrogen,
 the perturbation operator $V_i$
occurring in $ $ Eq.~(\ref{t}) is  given by
 $1/|{\mathbf r}_a-{\mathbf r}_b|-1/r_a$
(which is the part of the total Hamiltonian not diagonalized
by $|\Phi_{{\mathbf k}_{i}}\rangle$).
\begin{figure}
  \begin{center}
    \includegraphics[scale = 0.75]{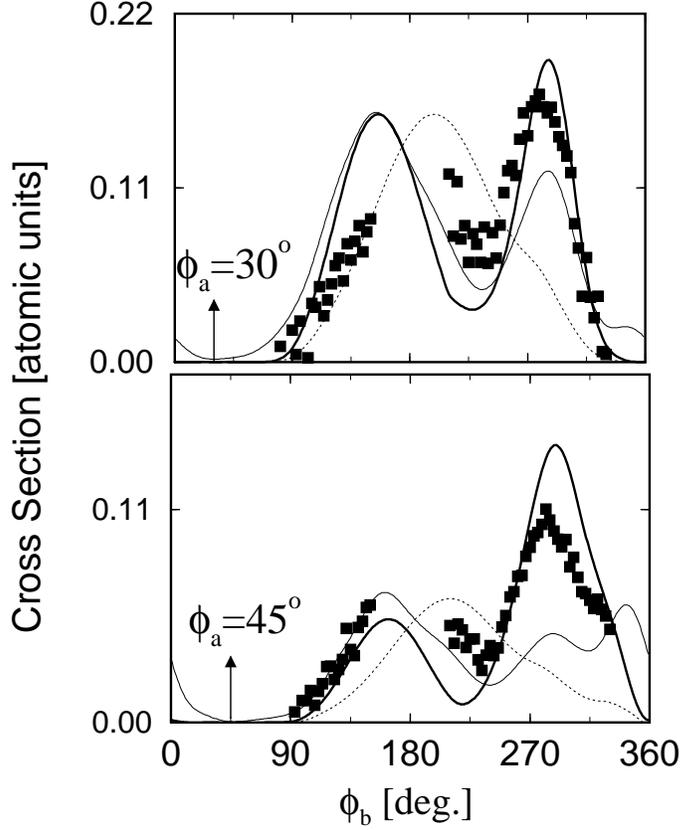}
    \caption{ The fully differential cross section for the
electron-impact ionization of atomic hydrogen in the co-planar,
symmetric energy-sharing geometry. The incident energy is $E_i=27.2\ eV$.
One ejected electron
is detected at a fixed    angle $\Phi_a$
with respect to the incident
 direction [
 $\Phi_a=30^{\rm o}$ (upper panel) and $\Phi_a=45^{\rm o}$ (lower pannel)].
 The angular distribution of
the other  emitted  electron
is measured. The emission angle of this
electron  with respect to the incident direction is denoted by $\Phi_b$.
Both electrons have the same energy, namely $ E_a=E_b=6.8\ eV$.
Full squares are experimental data 
Ref.\cite{jochp,jpbigor}.
 The solid thick lines show the predictions of the DS3C theory employing the
 matrix ${\cal A}$ (cf. Eq.(\ref{trans}))
 whereas
 the dotted curves indicate the results of the
3C theory, i.e. when using ${\cal A}\equiv 1$.
Thin solid lines are the full
numerical calculations using the convergent close coupling method (CCC). }
    \label{fig:fig4}
  \end{center}
\end{figure}
\begin{figure}
  \begin{center}
    \includegraphics[scale = 0.72]{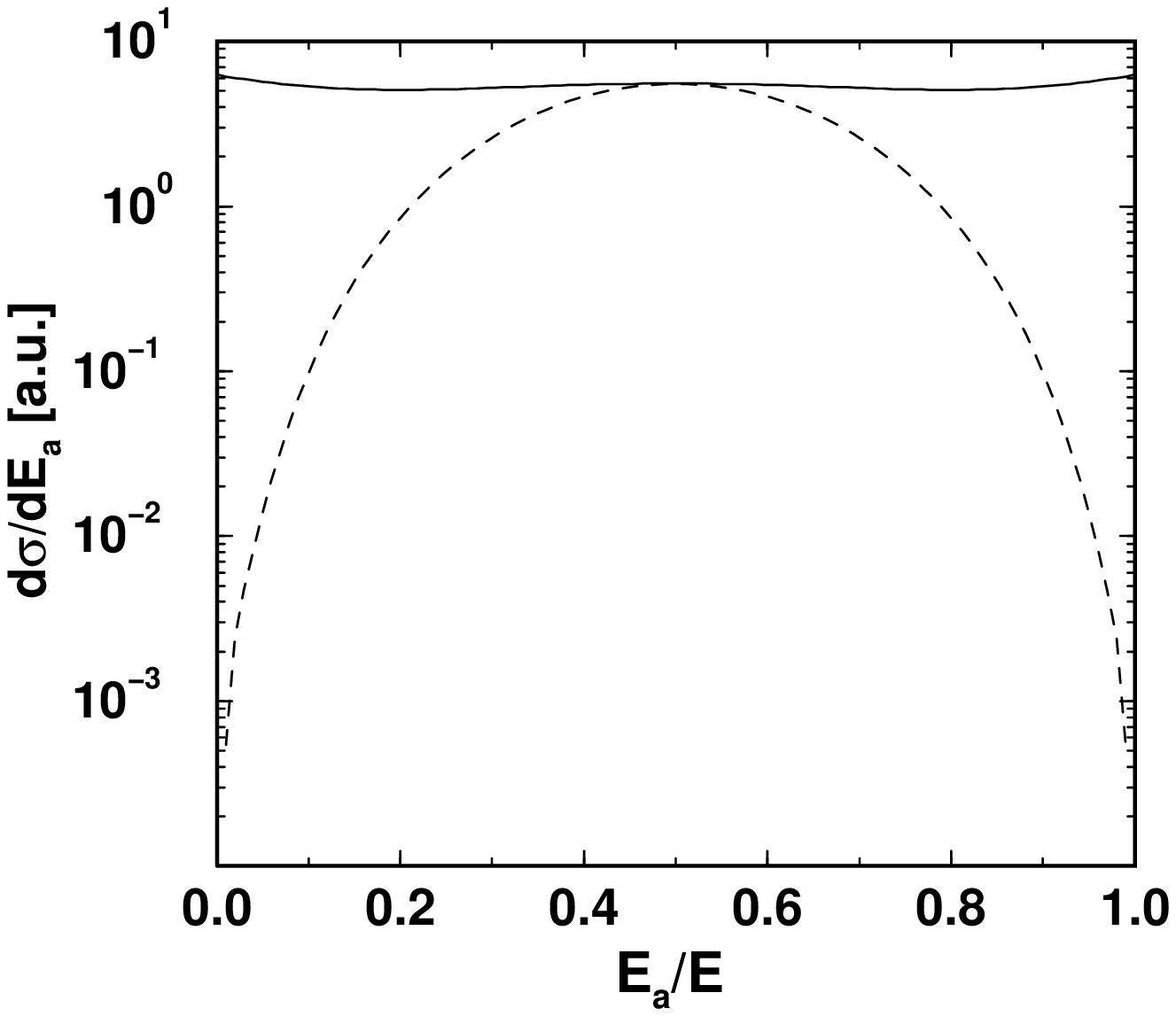}
    \caption{
The single differential cross section for the electron-impact
ionization of atomic hydrogen as function of the ratio $E_a/E$ where $E_a$ is
the energy of one of the final-state electrons and $E=E_a+E_b$ is the total excess
energy which is chosen as $E= 200 \ meV$.
 The use of  the $\Psi_{\rm DS3C}$ approximation
yields  the solid curve whereas the dashed curve  represents the results when
employing the 3C model (${\cal A}={\rm I}$). The 3C results
have been multiplied by a factor of $10^{13}$ for a better shape comparison. }
 \label{fig:fig3}
  \end{center}
\end{figure}
In what follows we choose the $x$ axis as the incident direction
$ \hat{\mathbf  k}_i$. The final state electrons are detected in a
coplanar geometry, i. e. ${\mathbf k}_i\cdot({\mathbf k}_a \times{\mathbf
 k}_b )=0$. The $z$ axis lies
along the direction perpendicular to the scattering plane,
i. e. parallel to $ \hat {\mathbf k}_a\times
 \hat{\mathbf k}_b$. The polar and azimuthal
angles of the vector ${\mathbf k}_a $ (${\mathbf k}_b $) are denoted by
$\theta_a,\phi_a$ ($\theta_b,\phi_b$),
 respectively. In the coplanar geometry considered
here the polar angles are fixed to $\theta_a=\pi/2=\theta_b$.
In  the calculation of the  DS3C model
we employ  the  approximation
(\ref{prop})  and use
 the product charges (\ref{zab}-\ref{zb}). If we
use the unit matrix for the transformation (\ref{trans}),  i.e.
if we assume ${\cal A}= {\rm I}$, the three-body wave function
reduces the eigenfunction of the asymptotic part $H_{\it par}$ of the
Hamiltonian without any coupling to the internal region. This
wave function is commonly known as the 3C wave function \cite{bbk,bbk91}.
 In addition we compare the results of the analytical methods
presented here with those of the convergent close coupling method (CCC).
This is a purely numerical method that attempts
at evaluating exactly the transition matrix elements fully numerically.
In Fig.1 the angular distribution of one of the
electrons is shown for two fixed angular positions of  the other
electron. The two electrons are ejected with
equal energies $E_a=E_b=6.8\, eV$. As clear from Fig.~1 the effect of the
coupling to the internal region is very important, since the
results of the 3C model that neglects the short-range dynamics are
at clear variance with the experiment.
The differences between the CCC method and the experiments are still the
subject of current research.
The main advantage of analytical methods is that they allow an insight into
the origin of the structures observed in the cross sections.
An extensive analysis underlying this statement has been carried out in
Ref.(\cite{jpbigor}) where the main peaks in Fig.~1 have been assigned to
certain sequence of collisions between the participating particles.

At higher incident energies the discrepancies between the DS3C and the
3C results disappear and both of those models (as well as the CCC calculations)
are in overall agreement with the experiments \cite{pra97}. From this
situation one can conclude that at higher energies the short-range
parts of the Hamiltonian ($H_{\it int}$ and $H_{\it mix}$)
are of less importance, for they have been neglected in
the 3C model whereas the DS3C theory accounts for them via the
dynamical screening (we note the asymptotic region is reached
for large $k_{ij}\xi_k$, i.e.~for large momenta the distance $\xi_k$ does not
 need to be
very large). \\
This means in physical terms that the
two electrons attains their asymptotic momenta swiftly without
much of  scattering from intermediate states whose behaviour is
determined mainly by  $H_{\it int}$ and $H_{\it mix}$.

Integrating over all emission angles of the two electrons we end up with
a single differential cross section depending on the energy of one of the
electrons. Since the energy of the other electron is then determined via
the energy-conservation law, the single differential cross section has to be
symmetric with respect to the point where both electrons have the same energy.
 Fig.~2  shows the results for the single differential cross sections
 as calculated within
 the DS3C method along with the calculations within  the 3C method.
 The excess energy is very
low ($200\, meV$). For small excess energies
 the Wannier theory, which relies on phase
space arguments, predicts a flat energy distribution between the electrons,
i.e. a flat single-differential cross section. This prediction has been
substantiated by full numerical calculations \cite{pont}. As seen in Fig.~2
the DS3C predicts a flat energy sharing between the electrons close to the
complete fragmentation threshold, in contrast to the 3C results which
are strongly peaked around the equal energy-sharing configuration.
This deviation of the 3C results from those of the Wannier theory
 is not surprising since in the Wannier approach  one expands the potential around
a saddle point (accounting for terms up to a fourth order)
and neglects higher order terms while the
kinetic energy is treated fully. In contrast the 3C model neglects the short-range
part of the kinetic energy. Obviously it is this part which is most important for the
Wannier mode and the resulting predictions.

Sampling over the energy sharing between the two electrons, i.e.
integrating the single differential cross section shown in Fig.~2,
one obtains the total cross section as function of the excess
energy $E=E_a+E_b$ (or equivalently as function of the incident energy $E_i$).
Close to the three-body break-up threshold the total cross section
$\sigma(E)$ for two continuum electrons receding from a charged ion
has been investigated by Wannier \cite{wan} using a
classical analysis. Wannier \cite{wan} pointed out that the
excess-energy functional dependence of
the total ionization cross section at
 the three-particle fragmentation threshold can be deduced
from the  volume of the
phase space available for double escape of the two electrons.
For the present case of  atomic hydrogen Wannier deduced
the threshold law $\sigma(E)\propto E^{1.127}$.
Since then an immense amount of theoretical and experimental studies
( e.g. \cite{peter1,peter2,cvej1,cvej2,read,rau71,klar,jim,kos,rost,lab,macek})
using quite different  approaches have been carried out which basically
confirm the Wannier-threshold law.\\
\begin{figure}
  \begin{center}
    \includegraphics[scale = 0.8]{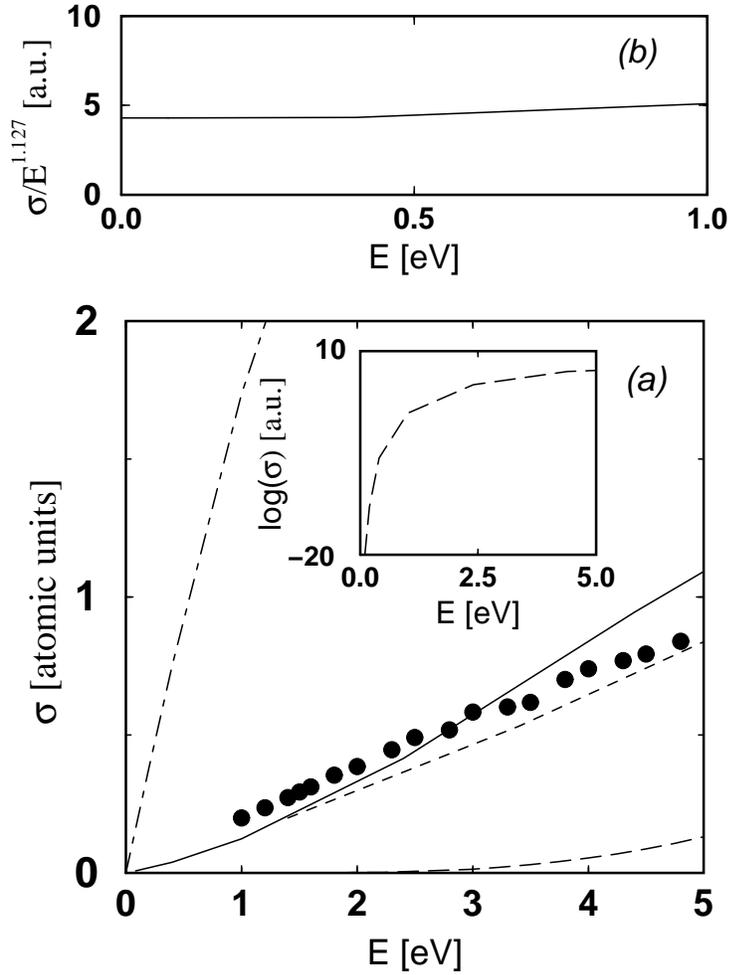}
    \caption{
The total
cross section $\sigma(E)$
 for the electron-impact ionization of atomic hydrogen as
a function of the excess energy $E$. The solid (long dashed) curve shows the results
for $\sigma(E)$ when treating  the two continuum electrons according
to the DS3C theory (3C model) whereas the dashed dotted curve denotes the
results of the independent Coulomb particles model (see text).
Results of the CCC method are also included (short dashed curve). Experimental
 data are due to Shah {\it et al.} \cite{shah}. The inset
in the panel (a) shows the results of the 3C theory on a logarithmic
scale.  In the upper panel (b) the quantity $\sigma(E)/E^{1.127}$  is depicted
as a function
of $E$ as evaluated using $\Psi_{\rm
DS3C}$.
}   \label{fig:fig1}
  \end{center}
\end{figure}
\begin{figure}
  \begin{center}
    \includegraphics[scale = 0.8]{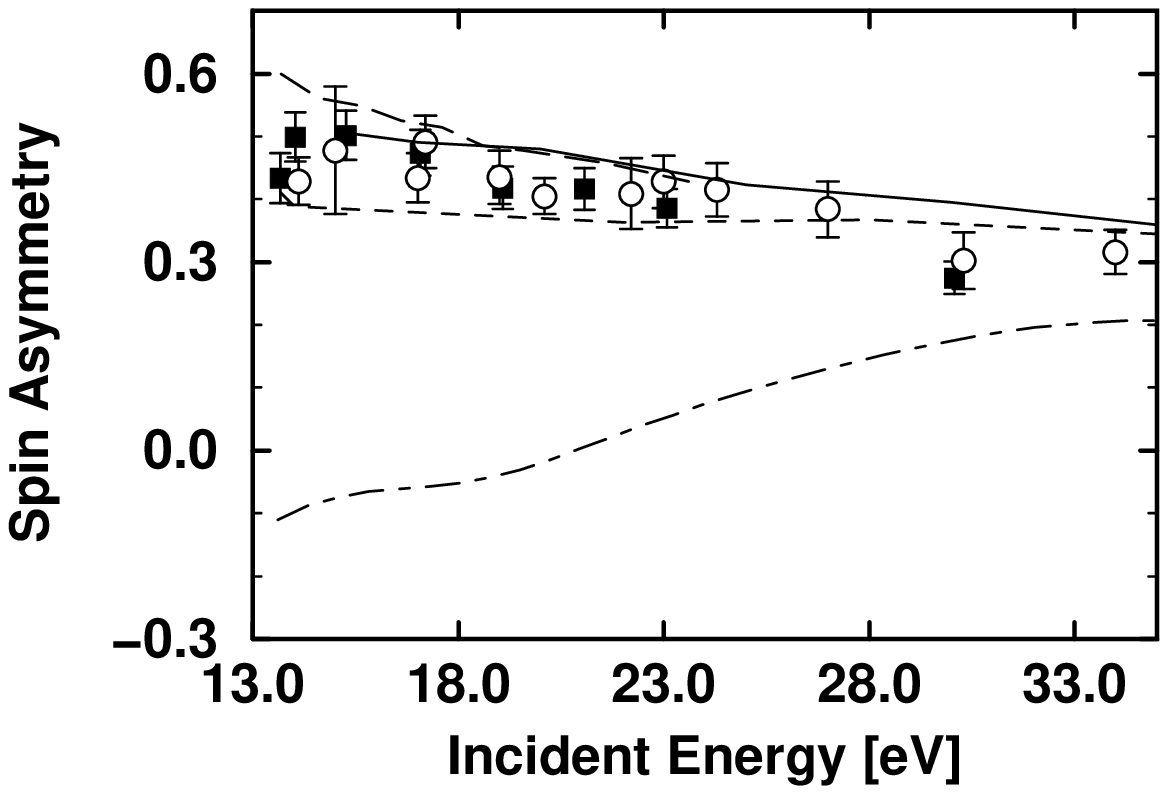}
    \caption{
The spin asymmetry [Eq.(\ref{asy})] in the total
ionization cross section $\sigma(E)$
  for the electron-impact ionization of atomic hydrogen.
Results of the hidden crossing theory \cite{macek}
(long dashed curve) and those of the  CCC method \cite{bray}
 (solid curve) are depicted. Treating the final
state according to the DS3C theory  (3C theory) yields for the spin asymmetry the results
shown by the  short-dashed (dot-dashed)  curve.
The experimental data are due to Fletcher {\it et al.} \cite{flet}
(full squares) and
Crowe {\it et al.} \cite{crow} (open circles).}
    \label{fig:fig2}
  \end{center}
\end{figure}
The Wannier treatment predicts the scaling behaviour of the
cross section $\sigma(E)$, but it does not provide any information about
the magnitude of $\sigma(E)$.
That the magnitude is a very sensitive quantity is
illustrated by the behaviour of the cross section in the independent Coulomb
particle model which is obtained
in our case by switching off the interaction between the
 two electrons in the final channel. In
this case the cross section reveals a    linear dependence on the excess energy,
$\sigma(E)\propto E$ \cite{zphysd}. Although the latter dependence of
$\sigma(E)$ does not deviate much
from the Wannier threshold law ($\sigma(E)\propto E^{1.127}$)
 the absolute value of
$\sigma(E)$ within the independent Coulomb particle model  is
largely overestimated [compare Fig.3].
If we employ the wave function $\Psi_{\rm DS3C}$,
with the dynamical product charges described in the preceding sections
we end up with  results in good accord with the
experimental measurements (cf.   Fig.3).
 The absolute magnitude
of the total cross section is satisfactorily
reproduced when  the DS3C model is employed.
  To examine the analytical behaviour of $\sigma(E)$ calculated using
$\Psi_{\rm DS3C}$ we plot in  Fig.~3(b) the quantity $\sigma(E)/E^{1.127}$.
According to the Wannier-threshold
 law the latter quantity $ $ should be a constant function of
$E$ and gives the absolute value of the cross section.
As seen in  Fig.~3(b) the
Wannier threshold law is in fact
 reproduced by the cross section results of the
 $\Psi_{\rm DS3C}$ within a range
of $E\in[0,0.5eV]$. For $E>0.5\ eV$ the analytical
dependence of $\sigma(E)$ evaluated with $\Psi_{\rm DS3C}$
 slowly deviates from the
Wannier threshold law.
 When using the 3C method for the description
of the two escaping electrons (${\cal A}={\rm I}$ in Eq.(\ref{trans}))
 we obtain an
analytical behaviour for $\sigma(E)$  which is
not compatible with the Wannier theory.
The  absolute value
for the total cross section is as well not reproduced by the 3C model,
for the reasons discussed above.
 Also included in Fig.3 are the results of the
convergent-close coupling method, CCC, \cite{bray}. The results of the
CCC are in  good agreement with the experimental $\sigma(E)$ for higher
energies \cite{bray},
 however, close to threshold the evaluation of $\sigma(E)$ is
limited by the computational resources as an ever increasing number of pseudo states is
needed to achieve convergence.\\
In addition to the
magnitude of the cross section, the spin asymmetry, $A$, offers
a further way of probing the dynamical properties of the electron-impact
ionization of atomic systems. The spin asymmetry $A$ is defined as
\begin{eqnarray}
A(E):=\frac{\sigma^s(E)-\sigma^t(E)}{\sigma^s(E) + 3 \sigma^t(E)}
,\label{asy}\end{eqnarray}
where $\sigma^s$ and $\sigma^t$ are the total ionization cross sections
for singlet and triplet scattering, respectively.
 The Wannier theory for threshold ionization predicts a
constant value
of $A$ with increasing excess energy but provides no information
on the numerical value of $A$ \cite{green}. Measurements of $A$ at threshold
 reveals a slightly
positive slope of the
 spin asymmetry with increasing excess energies \cite{guo}.
In Fig.4
the results for $A$ are shown in the case where the
two-electron continuum final state is described by the 3C theory
and by  $\Psi_{\rm DS3C}$. Also depicted in Fig.4 are the results of the
CCC approach \cite{bray}
 and the method using hidden-crossing
theory \cite{macek}.
 Although all theories,  except for the 3C model,
are in reasonable agreement with experimental finding the positive
slope of $A$ at threshold is not reproduced.

Neglect of the short-ranged part of the Hamiltonian $H_{\it in}$ and
$H_{\it mix}$, i.e. using the 3C model,  results
in a completely wrong behaviour of the calculated
spin asymmetry.
 With increasing excess energy the inner  region of the Hilbert
space becomes of less importance for the present reaction and
the results of the 3C method become more and
 more
in better agreement with the experimental data.

We note here that since the spin asymmetry is a ratio
of cross sections it is expected that the spin asymmetry
is rather sensitive to the detailed of the radial part of the
wave functions. From the agreement between the experiment and
the DS3C theory observed in Fig.~4 we conclude that the
radial part of
 $\Psi_{\rm DS3C}$ is well behaved at lower excess energies and that
the short-range parts of the total Hamiltonian $H_{\it int}$ and
$H_{\it mix}$ plays a dominate role at lower  energies,
as far as the value and the behaviour of the spin asymmetry are
concerned.
%
%
%
%
%
%
\section{Correlated states of  N charged particles}
In the preceding sections we considered the two and three-body continuum
spectrum. Unfortunately,
the curvilinear  coordinate system (\ref{cor},\ref{cor1}) used for the
three-body problem does not have a straightforward generalization to the
$N$ body case. Therefore we will treat the problem of 
$N$ charged particles at energies  above the complete
fragmentation threshold within a reference frame spanned by  a set
of Jacobi coordinates.
The problem becomes more transparent if we consider
 $N-1$ particles with equal masses
(and with charges $z_j,\ j\in[1,N-1]$) that move
in the Coulomb field of a residual massive charge $z$.
The mass $M$ of the charge  $z$ is  assumed to be
much larger than $m$ ($M\gg m$).  Neglecting terms of the order
$m/M$ the centre-of-mass system and the laboratory frame of reference
can be chosen to be identical. 
The non-relativistic
 time-independent Schr\"odinger equation of the $N$-body
system
can then be formulated in the relative-coordinate representation as
\begin{eqnarray}
\left[ H_0 + \sum^N_{j=1}\frac{zz_j}{r_{j}}\ +
\sum^N_{\stackrel{i,j}{j>i=1}}\frac{z_iz_j}{r_{ij}}
-E\right]
 \ \Psi({\bf r}_1,\cdots,{\bf r}_{N}) =0\ 
\label{schra}\ \end{eqnarray}
where ${\bf r}_{j}$ is
 the position of particle  $j$  with respect to the residual 
charge $z$ and ${\bf r}_{ij}:={\bf r}_{i}-{\bf r}_{j}$
 denotes the relative coordinate between
 particles  $i$  and  $j$ . The kinetic energy operator $H_0$ has  the form
(in the limit $m/M\to 0$)
 $H_0 =-\sum_{\ell=1}^N \Delta_\ell/2m$ where $\Delta_\ell$ is the
Laplacian with respect to the coordinate ${\bf r}_{\ell}$.
 We note here that for a system of general
masses the problem is complicated by
an additional mass-polarization term
  which arises in Eq.\  (\ref{schra}). Upon
introduction of
$N$-body
 Jacobi coordinates, $H_0$ becomes diagonal, however, the potential terms
acquire a much more
complex form. \\
Assuming the continuum particles
 to escape with relative asymptotic momenta
${\bf k}_{j}$ (with respect to the charge $z$)
it has been suggested in Ref.\  \cite{red}, due to
unpublished work by Redmond, that for
large interparticle
 distances  the wave function
 $\Psi({\bf r}_{1},\cdots,{\bf r}_{N})$ takes on
the form
\begin{eqnarray}
 \smash{\lim_{\stackrel{r_{lm}\to\infty}{r_{n}\to\infty}}}
\Psi({\bf r}_{1},\cdots,{\bf r}_{N})\to
   (2 \pi )^{- 3N/2}
  \prod_{s=1}^N\bar\xi_s({\bf r}_{s})\psi_s({\bf r}_{s})
\  \prod_{\stackrel{i,j=1}{j>i}}^N\psi_{ij}({\bf r}_{ij})\ ,
\label{reda}\end{eqnarray}
where the functions $\bar\xi_j({\bf r}_{j}),\
 \psi_j({\bf r}_{j}),\ \psi_{ij}({\bf r}_{ij})$
are defined as
\begin{eqnarray}
\bar\xi_j({\bf r}_{j})&:=&\exp(i{\bf k}_{j}\cdot{\bf r}_{j})\\
\psi_j({\bf r}_{j}) &:=&
 \exp\left[\mp i{\alpha}_{j}\ln(k_{j}\ r_{j}\pm
{\bf k}_{j}\cdot{\bf r}_{j})\frac{ }{}\right]\ ,\\
\psi_{ij}({\bf r}_{ij}) &:=&
 \exp\left[\mp i{\alpha}_{ij}\ln(k_{ij}\ r_{ij}\pm
{\bf k}_{ij}\cdot{\bf r}_{ij})\frac{ }{}\right]\ .
\label{phases}\ \end{eqnarray}
The $+$  and $-$ signs refer to
 outgoing and incoming
boundary conditions, respectively, and
${\bf k}_{ij}$ is the momentum conjugate to
${\bf r}_{ij}$, i.\ e.\
 $ {\bf k}_{ij}:=({\bf k}_{i}-{\bf k}_{j})/2$.
 The Sommerfeld-parameters
 ${\alpha}_{j},
{\alpha}_{ij}$ are given by
\begin{eqnarray}
{\alpha}_{ij}=\frac{z_iz_j}{v_{ij}},\hspace*{1cm}
{\alpha}_{j}=\frac{zz_j}{v_{j}}
.\label{somp}\ \end{eqnarray}
In Eq.\  (\ref{somp}) $v_{j}$
 denotes the velocity of particle  $j$  relative to
the residual charge whereas ${\bf v}_{ij}:=\bf{v}_i -\bf{v}_{j}$.
In this work we restrict the considerations to outgoing-wave boundary
conditions. The treatment of incoming-wave boundary conditions runs
along the same lines.
The total energy of the system $E$ is given by
\begin{eqnarray}
E=\sum_{l=1}^NE_l,\
 \mbox{ where } \ E_l=\frac{k^2_l}{2m}.\label{E}\end{eqnarray}
To derive asymptotic
 scattering states in the limit of large inter-particle
separations and their propagations to finite distances we assume for
$\Psi({\bf r}_{1},\cdots,{\bf r}_{N})$  the ansatz
\begin{eqnarray}
\Psi({\bf r}_{1},\cdots,{\bf
 r}_{N})={\cal N} \Phi_I({\bf r}_{1},\cdots,{\bf r}_{N})\Phi_{II}
({\bf r}_{1},\cdots,{\bf r}_{N})\chi({\bf r}_{1},\cdots,{\bf r}_{N})
\label{an}\end{eqnarray} where $\Phi_I,\ \Phi_{II}$ are appropriately
chosen functions, ${\cal N}$ is a normalization constant
and  $\chi({\bf r}_{1},\cdots,{\bf r}_{N})$
 is a function of an arbitrary form.
The function $\Phi_I$ is chosen to describe the motion of $N$-independent
Coulomb particles moving in the field of the charge $z$ at the total
energy $E$, i.\ e.\ \ $\Phi_I$ is determined by the differential equation
\begin{eqnarray}
\left(H_0 + \sum_{j=1}^N\frac{zz_j}{r_j} -E\right)
\Phi_I({\bf r}_{1},\cdots,{\bf r}_{N})=0.
\label{phi1}\end{eqnarray}
Since we are interested in
 scattering solutions with outgoing-wave boundary
conditions which describe $N$-particles escaping with asymptotic
momenta ${\bf k}_{j}, \ j\in[1,N]$, it is appropriate to factor 
out the plane-wave part and write for $\Phi_I$
\begin{eqnarray}
\Phi_I({\bf r}_{1},\cdots,{\bf r}_{N})=
\overline{\Phi}_I({\bf r}_{1},\cdots,{\bf
 r}_{N})\prod_{j=1}^N\bar\xi_j({\bf r}_{j}).
\label{phi1o}\end{eqnarray}
Upon substitution of the ansatz (\ref{phi1o}) into
Eq.\  (\ref{phi1}) it is readily  concluded that Eq.\  (\ref{phi1})
is completely separable  and the regular solution
$\Phi_I$ can be written in closed form
\begin{eqnarray}
\Phi_I({\bf r}_{1},\cdots,{\bf
 r}_{N})=\prod_{j=1}^N\bar\xi_j({\bf r}_{j})\varphi_j({\bf r}_{j})
\label{phi1e}\end{eqnarray}
where  $\varphi_j({\bf r}_{j})$ 
 is a confluent-hypergeometric function in the
notation of Ref.\  \cite{abra}
\begin{eqnarray}
\varphi_j({\bf r}_{j})=
 \ _1F_1[\alpha_j,1,-i(k_jr_j+{\bf k}_{j}\cdot{\bf r}_{j})].
\end{eqnarray}
The function $\Phi_I$ describes the motion of the continuum particles in the
extreme case of very strong coupling to the residual ion, i.\ e.\  $|zz_j|\gg
|z_jz_i|;\ \forall\  i,j\in[1,N]$. In order to incorporate the other extreme
case of strong correlations among 
  the continuum particles ($
|z_jz_i|\gg |zz_j|;\ \forall\ 
 i,j\in[1,N]$) we choose $\Phi_{II}$ to possess the  form
\begin{eqnarray}
\Phi_{II}({\bf r}_{1},\cdots,{\bf
 r}_{N})=\overline{\Phi}_{II}({\bf r}_{1},\cdots,{\bf r}_{N})
\prod_{j=1}^N\bar\xi_j({\bf r}_{j})
\label{phi2}\end{eqnarray}
with 
\begin{eqnarray}
\overline{\Phi}_{II}({\bf r}_{1},\cdots,{\bf r}_{N})
:=\prod_{j>i=1}^N\varphi_{ij}({\bf r}_{ij})
\label{phi2o}\end{eqnarray}
where $\varphi_{ij}({\bf r}_{ij})
:=\ _1F_1[\alpha_{ij},1,-i(k_{ij}r_{ij}+{\bf k}_{ij}\cdot{\bf r}_{ij})]$.
It is straightforward to show that  the expression  
$\varphi_{ij}({\bf r}_{ij})\prod_{l=1}^N\bar\xi_l({\bf r}_{l})$ solves for the
Schr\"odinger Eq.\  (\ref{schra}) in the case of extreme correlations
between particle  $i$  and particle  $j$ , i.\ e.\ \ $|zz_l|\ll|z_iz_j|\gg
|z_mz_n|,\ \forall\ l,m,n\neq i,j$.
 In terms of differential equations this means
\begin{eqnarray}
\left(H_0 + \frac{z_iz_j}{r_{ij}}
 -E\right)\varphi_{ij}({\bf r}_{ij})\prod_{j=1}^N\bar\xi_j({\bf r}_{j})=0.
\label{phi12}\end{eqnarray}
It should be stressed, however, 
that the function (\ref{phi2}) does not solve for
Eq.\  (\ref{schra}) in case of weak coupling to the residual ion
 ($z\to 0$), but otherwise comparable strength of correlations between the
continuum particles. This is due to the fact
that two-body subsystems
 formed by the continuum particles are coupled to each
other. To derive an expression for this coupling term we note first that
\begin{eqnarray}
\Delta_m\overline{\Phi}_{II}&=&
\sum_{l=1}^{m-1}\Delta_m\varphi_{lm}\prod_{\stackrel{j>i}{i\neq
               l}}^N\varphi_{ij}\ + \
\sum_{n=m+1}^N\Delta_m\varphi_{mn}\prod_{\stackrel{j>i}{j\neq
               n}}^N\varphi_{ij}
+ A_m, \hspace*{1cm}m\in[1,N] \label{dell}\end{eqnarray} 
where the differential operator $A_m$ has the form
\begin{eqnarray}
A_m &=&
 2 \sum_{l=1}^{m-1}\left[( {\bf \nabla}_m\varphi_{lm})\cdot
( \sum_{n=m+1}^{N}
{\bf \nabla}_m\varphi_{mn})\right]\prod_{\stackrel{j>i}{j\neq
               n, i\neq l}}^N\varphi_{ij}\nonumber\\
&& + \sum_{l=1}^{m-1}\left[ ({\bf \nabla}_m\varphi_{lm})\cdot
(\sum_{l\neq s= 1}^{m-1} {\bf \nabla}_m\varphi_{sm}
)\right]\prod_{\stackrel{j>i}{s\neq i\neq l}
               }^N\varphi_{ij}\nonumber\\
&& +  \sum_{n=m+1}^{N}\left[ ({\bf \nabla}_m\varphi_{mn})\cdot(
\sum_{\stackrel{t=m+1}{t\neq n}}^{N} {\bf \nabla}_m\varphi_{mt}
)\right]\prod_{\stackrel{j>i}{j\neq t\neq n}
               }^N\varphi_{ij}
\hspace*{1cm}m\in[1,N].
\label{Am}\end{eqnarray}
To obtain the differential operator which couples the
 two-body subsystems
in absence of the charge $z$ we  neglect in
(\ref{schra}) the interactions between the residual charge and
 the continuum particles ($z=0$) and substitute the function
 (\ref{phi2}) into
Eq.\  (\ref{schra}). Making use of the relation (\ref{dell})
 it is straightforward,
however cumbersome, to show that the coupling
 term which prevent separability
has the form
\begin{eqnarray}
A=\sum_{m=1}^N A_m.
\label{A}\end{eqnarray}
Eqs.\ (\ref{Am},\ref{A}) warrant commented upon:
The term $A_m$ is a mixing operator. It couples an
individual two-body
subsystem formed by two continuum particles to all other two-body
subsystems formed by the continuum particles in absence of the residual
ion. Hence it is clear that
 all the terms in the sum (\ref{Am})
 vanishes for the case of three-body system
since in this case only one two-body
 system does exist in the field of the residual
charge. The second remark concerns the structure
 of $A_m$ and hence $A$. From Eq.\  (\ref{dell})
it is evident
 that the remainder term (\ref{Am}) is part of the kinetic energy operator.
Thus it is expected that, under certain circumstances,
 this term has a finite range which indicates
 that asymptotic separability, in the sense specified below, does exist
 for many-body continuum Coulomb systems. In fact, as the functional form
of $\varphi_{ij}({\bf r}_{ij})$
 is known the term $A$ can be calculated explicitely
which will be done below.\\
Now with $\Phi_I$ and $\Phi_{II}$ 
 have been determined, the exact wave function
(\ref{an})  is given by the expression $\chi({\bf r}_{1},\cdots,{\bf r}_{N})$. 
Upon substitution of the expressions (\ref{phi2},\ref{phi1e}) into the
ansatz
(\ref{an}) and inserting
in the Schr\"odinger equation (\ref{schra}) a differential equation
for the determination of $\chi({\bf r}_{1},\cdots,{\bf r}_{N})$ is derived
\begin{eqnarray}
&&\left\{H_0 - \frac{A}{\overline{\Phi}_{II}} -
\sum_{\ell=1}^N\left[({\bf \nabla}_{\ell}\ln\Phi_I\ + \
    {\bf \nabla}_{\ell}\ln\Phi_{II})\cdot{\bf
 \nabla}_{\ell}\frac{}{}\right.\right. \nonumber\\
&&\left.\left. + ({\bf \nabla}_{\ell}\ln\Phi_I)\cdot({\bf
 \nabla}_{\ell}\ln\Phi_{II})\frac{}{}\right]
+ E\right\}\chi({\bf r}_{1},\cdots,{\bf r}_{N})=0
\ .\label{chi}\end{eqnarray}
From the derivation of the functions $\Phi_{I}$ and $\Phi_{II}$ 
[Eqs.\ (\ref{phi1},\ref{phi2})] it is clear that all long-range two-body
Coulomb interactions
have been already diagonalized by $\Phi_{I}$ and $\Phi_{II}$ because
the total potential is exactly treated by these wave functions. Hence, the
function $\chi$, to be determined here, contains information
on many-body couplings, which are, under certain conditions (see below), 
of  finite range.
 To explicitely show that,
 and due to 
flux arguments we write the function $\chi$ in the form
\begin{eqnarray}
\chi({\bf r}_{1},\cdots,{\bf r}_{N})=
 \prod_{j=1}^N{\bar\xi}^*({\bf r}_{j})[1-f({\bf r}_{1},\cdots,{\bf r}_{N})]
\label{chan}\end{eqnarray}
where $f({\bf r}_{1},\cdots,{\bf r}_{N})$
 is a function of an arbitrary structure.
Inserting the form (\ref{chan}) into Eq.\  (\ref{chi}) we arrive, after
much differential analysis, at the inhomogeneous differential equation
\begin{eqnarray}
\left\{ H_0 - \sum_{\ell=1}^N\left[{\bf \nabla}_{\ell}(\ln\Phi_I\ + \
    \ln\Phi_{II}) +
 i{\bf k}_{\ell}\right]\cdot{\bf \nabla}_{\ell}\right\}f + {\cal R}(1-f)=0
\label{f}\end{eqnarray}
where the inhomogeneous term ${\cal R}$ is given by
\begin{eqnarray}
{\cal R}:=\sum_{m=1}^N&&\left\{
 ({\bf \nabla}_{m}\ln\overline{\Phi}_{I})\cdot({\bf
 \nabla}_{m}\ln\overline{\Phi}_{II})
   + \sum_{l=1}^{m-1}\sum_{p=m+1}^N({\bf \nabla}_m\ln\varphi_{lm})
\cdot ({\bf \nabla}_m\ln\varphi_{mp})\right.\nonumber\\
&&\left. +\frac{1}{2}
\sum_{l=1}^{m-1}\sum_{s\neq l}^{m-1}({\bf \nabla}_m\ln\varphi_{lm})
\cdot ({\bf \nabla}_m\ln\varphi_{sm})\right.
\nonumber\\
&&+\left.\frac{1}{2}
\sum_{n=m+1}^{N}\sum_{n\neq q=m+1}^N({\bf \nabla}_m\ln\varphi_{mn})
\cdot ({\bf \nabla}_m\ln\varphi_{mq})
               \right\}.
\label{R}\end{eqnarray} 
It is the inhomogeneous term ${\cal R}$
 which contains the coupling between all individual
two-particle subsystems. For example the first term in Eq.\ 
 (\ref{R}) describes
the coupling of a two-body subsystems formed by particles  $i$  and  $j$ 
to all 
two-body subsystems  formed by
 the individual continuum particles  and the residual ion.
The second term originates from (\ref{A}) and,
 as explained above, is a measure for the coupling among  
two-body subsystems of the continuum particles
 (in absence of $z$).
To these couplings to be negligible the norm of the
term ${\cal R}$ must be small. For get some insight into the functional form
of ${\cal R}$, given by (\ref{R}), we note
that
\begin{eqnarray}
{\bf \nabla}_{\ell}\ln\overline{\Phi}_{I}=\alpha_{\ell}k_{\ell}\
 {\bf F}_{\ell}({\bf r}_{\ell})
\label{f12}\end{eqnarray}
where
\begin{eqnarray}
{\bf F}_{\ell}({\bf r}_{\ell})&:=&
{   \frac{
\mbox{} _1F_1 \left[ 1+i\alpha_{\ell},\
 2,\ -i(k_{\ell}\ r_{\ell}+{{\bf k}_{\ell}} \cdot
{{\bf r}_{\ell}})\
\right] } {
 _1F_1 \left[ i\alpha_{\ell},\
 1,\ -i(k_{\ell}\ r_{\ell}+{{\bf k}_{\ell}} \cdot {{\bf r}_{\ell}})\
\right]
  }\ ({\bf \hat k}_{\ell}+{\bf \hat r}_{\ell}) }
\label{bruch1}\ .\ \end{eqnarray}
In addition we remark that
\begin{eqnarray}
{\bf \nabla}_{m}\ln\overline{\Phi}_{II} &=&
 \sum_{n=m+1}^N {\bf \nabla}_m\ln\varphi_{mn}
+ \sum_{l=1}^{m-1} {\bf
 \nabla}_m\ln\varphi_{lm}\label{hlp1}\nonumber\\
&=& 
\sum_{n=m+1}^N \alpha_{mn}k_{mn}{\bf F}_{mn}({\bf r}_{mn})
-\sum_{l=1}^{m-1}\alpha_{lm}k_{lm}{\bf
 F}_{lm}({\bf r}_{lm})\label{phi2nab}
\end{eqnarray}
where
\begin{eqnarray}
{\bf F}_{ij}({\bf r}_{ij}) &:=&
{   \frac{
\mbox{} _1F_1 \left[ 1+i\alpha_{ij},\
 2,\ -i(k_{ij}\ r_{ij}+{{\bf k}_{ij}} \cdot
{{\bf r}_{ij}})\
\right]
                              } {
 _1F_1 \left[ i\alpha_{ij},\
 1,\ -i(k_{ij}\ r_{ij}+{{\bf k}_{ij}} \cdot {{\bf r}_{ij}})\
\right]
  }\ ({\bf \hat k}_{ij}+{\bf \hat r}_{ij}) }
\label{bruch12}\ .\ \end{eqnarray}
Thus, the behavior of the coupling term ${\cal R}$ is controlled by
the generalized functions ${\bf F}_{ij}({\bf r}_{ij}),\
 {\bf F}_{l}({\bf r}_{l})$
since Eq.\  (\ref{R}) can be written in the form
\begin{eqnarray}
{\cal R}:=\sum_{m=1}^N&&\left\{ \alpha_{m}k_{m}{\bf
F}_{m}({\bf r}_{m})\cdot\left[
\sum_{n=m+1}^N \alpha_{m n}k_{m n}{\bf F}_{m n}({\bf r}_{m n})
-\sum_{s=1}^{m -1}\alpha_{s m}k_{s m}{\bf F}_{s m}
({\bf r}_{s m})\right]\right.\nonumber\\
&-& \sum_{l=1}^{m-1}\sum_{p=m+1}^N \alpha_{lm}\alpha_{mp}k_{lm}k_{mp}
{\bf F}_{lm}\cdot {\bf F}_{mp} +\frac{1}{2} \sum_{l=1}^{m-1}\sum_{s\neq
l}^{m-1} \alpha_{lm}\alpha_{sm}k_{lm}k_{sm}
{\bf F}_{lm}\cdot {\bf F}_{sm}\nonumber\\
&& +\left.   \frac{1}{2}
\sum_{n=m+1}^{N}\sum_{n\neq q=m+1}^N \alpha_{mn}\alpha_{mq}k_{mn}k_{mq}
{\bf F}_{mn}\cdot {\bf F}_{mq} \right\}
\label{Rend}.\end{eqnarray}
The simplest approximation is to neglect the term ${\cal R}$ altogether.
In this case the function $f=0$ solves for
 the equation (\ref{f}). Then,
the solution of Eq.\  (\ref{schra}) takes on the approximate form
\begin{eqnarray}
\Psi({\bf r}_1,\cdots,{\bf r}_{N})\approx {\cal  N}
 \prod_{m>l,j=1}^N\bar\xi_j({\bf r}_{j})
\varphi_j({\bf r}_{j})\varphi_{lm}({\bf r}_{lm}).
\label{n3c}\end{eqnarray}
Thus, the justification of the approximation (\ref{n3c}) reduces to the
validity of neglecting the inhomogeneous term
 (\ref{Rend}). One region in which
this term can be disregarded is the asymptotic region of large inter-particle
separations. This is  immediately deduced from the asymptotic behavior of the 
generalized functions ${\bf F}_{ij}({\bf r}_{ij}),\
 {\bf F}_{l}({\bf r}_{l})$ which
dictate the asymptotic properties
 of ${\cal R}$, as readily concluded from
Eq.\  (\ref{Rend}). From the asymptotic expansion of the hypergeometric
functions \cite{abra}  we infer that
\begin{eqnarray}
\lim_{r_{ij}\to\infty}|{\bf F}_{ij}({\bf r}_{ij})|\ \to \
 \left|\ \frac{ {\bf \hat k}_{ij}+ {\bf \hat r}_{ij}}{{\bf k}_{ij}
\cdot( {\bf \hat k}_{ij}
+ {\bf \hat r}_{ij})\ r_{ij} }\ \right|\
 +\  {\cal O}\left( |k_{ij}\ r_{ij}+{\bf k}_{ij}\cdot{\bf r}_{ij}
|^{-2}\right) \ .
\label{basy}\ \end{eqnarray}
A asymptotic relation
 similar to Eq.\  (\ref{basy}) holds for ${\bf F}_{l}({\bf r}_{l})$.
It should be
 noted that the functions
 ${\bf F}_{ij}({\bf r}_{ij}),\ {\bf F}_{l}({\bf r}_{l})$
have to be considered in a 
distributive (operator) sense which means that,
asymptotically, only
terms of ${\bf F}_{ij},
 {\bf F}_{l}$ which fall off faster than the Coulomb potentials
can be disregarded. Since ${\cal R}$ is essentially  a sum of products
of ${\bf F}_{ij}, {\bf F}_{l}$ the expression ${\cal R}$ is of finite range,
in the sense that it
 diminishes faster than the Coulomb potential in the asymptotic regime,
 only in the case where all particles are far apart from each other, i.\ e.\ 
\begin{eqnarray}
\lim_{\stackrel{{r_{ij}\to\infty}}{r_l\to \infty}}{\cal R}\ \to\ {\cal O}
\left( |k_{ij}\ r_{ij}+{\bf k}_{ij}\cdot {\bf r}_{ij}|^{-2},
|k_{l}\ r_{l}+{\bf k}_{l}\cdot {\bf r}_{l}|^{-2}\right) \
\hspace*{0.3cm}\forall\hspace*{0.2cm}j>i,l\in[1,N].
\label{rasy}
\ \end{eqnarray}
Therefore, in the limit (\ref{rasy}), the term ${\cal R}$ can be asymptotically
neglected and the approximation (\ref{n3c}) is justified. In fact, it is
straightforward to show that the wave function (\ref{n3c}) tends to
the asymptotic form (\ref{reda}) in the limit of large inter-particle separations
which proves the assumption made in Ref.\  \cite{red}. However, if two particles
are close together, regardless of
 whether all other particles are well separated,
the coupling term is  of infinite range, as seen
 from Eqs.\ (\ref{basy},\ref{Rend}). In this case
 the relation (\ref{rasy}) does not hold.  Consequently, the wave function 
(\ref{n3c}) is not an exact asymptotic eigenfunction of the total Hamiltonian
in this limit. It is important to note that the limit
Eq.\ (\ref{rasy}) is energy dependent. With increasing
momenta of the escaping particles the asymptotic region, i.\ e.\ the
limit Eq.\ (\ref{rasy}), is reached faster.
In other words, at a certain inter-particle separations,
the remainder term ${\cal R}$, which
has been neglected to arrive at the approximate form
(\ref{n3c}), diminishes with increasing velocities of the emerging
particles. In this sense the approximation leading to the wave function
(\ref{n3c}) is a high energy approximation.
\subsection{The two-body cusp conditions}
In the preceding section it has been shown that the approximation
(\ref{n3c}) is, to leading order, exact  for large particles separation.
In addition, 
 it is concluded below that this function exhibits a behavior compatible with
equation (\ref{schra}) at all two-body coalescence points
$r_{ij}\to 0, r_{l}\to 0,\ j>i,l\in[1,N]$. To guarantee regular
behavior of the wave function at these collision points, at which
the corresponding Coulomb two-body potential is divergent, the solution
$\Psi({\bf r}_{1},\cdots,{\bf r}_{N})$ of Eq.\  (\ref{schra})
 must satisfy the Kato cusp conditions \cite{kat57,mye91}
(provided the solution does not vanish at these points). At a collision
point $r_{i}\to 0$ these conditions  are
\begin{eqnarray}
\left[\frac{\partial\ \tilde{\Psi}({\bf r}_{1},\cdots,{\bf r}_{N})}{\partial\
r_{i}}\right]_{r_{i}=0}&=&k_{i}\alpha_{i}
\Psi({\bf r}_{1},\cdots,r_{i}=0,\cdots,{\bf
 r}_{N})\hspace*{0.5cm}\nonumber\\
&&\forall\hspace*{0.2cm}(r_i/r_j)\to 0, (r_i/r_{lm})\to 0;\ m>l,\ i\neq
j\in[1,N]
\ .\label{cuspb}\ \end{eqnarray}
The quantity $\tilde{\Psi}({\bf r}_{1},\cdots,{\bf r}_{N})$ is the wave function
${\Psi}({\bf r}_{1},\cdots,{\bf r}_{N})$
averaged over a sphere of small radius $r_{\delta}\ll 1$ around
the singularity $r_{i}=0$. A relation similar to
Eq.\  (\ref{cuspb}) holds in the case of the coalescence
points $r_{ij}\to 0$. To prove  that the wave function
(\ref{n3c}) does satisfy the conditions (\ref{cuspb})  we linearize
$\Psi({\bf r}_{1},\cdots,{\bf r}_{N})$ around $r_i=0$ and
average over a sphere of small radius $r_{\delta}\ll 1$ 
to arrive at
\begin{eqnarray}
\tilde{\Psi}({\bf r}_{1},\cdots,{\bf
 r}_{N})&=&{\cal N}\ D({\bf r}_{i})\prod_{
\stackrel{i\neq
j=1}{l>m}}^N\bar\xi_j\varphi_j({\bf r}_{j})\varphi_{lm}({\bf r}_{lm}),\
\epsilon_{ilm}\neq 0
\label{cusp3}\end{eqnarray}
where
\begin{eqnarray}
D({\bf r}_{i}) &=& \frac{2\pi}{4\pi
r_{\delta}^2}\int_{-1}^{1}\!r_{\delta}^2\
d\cos\theta\ \left[\ \frac{}{}1+i k_{i}\cos\theta+{\alpha}_{i}k_{i}\
r_{i}(1 +
\cos\theta)
\ \right]
\nonumber\\
&=&1+{\alpha}_{i}\ k_{i}\ r_{i}
\ . \label{cusp2}
\ \end{eqnarray}
To arrive at Eq.\  (\ref{cusp2}) one takes the $z-$axes as
${\bf k}_{i}$ and define
 $\cos\theta= {\bf \hat k}_{i}\cdot {\bf \hat r}_{i}$.
From  Eqs.\ (\ref{cusp2},\ref{cusp3}) it is obvious that
\begin{eqnarray}
\left[\frac{\partial\ \tilde{\Psi}
({\bf r}_{1},\cdots,{\bf r}_{N})}{\partial\
r_{i}}\right]_{r_{i}=0}
&=& \alpha_i k_i\ {\cal N}\prod_{
\stackrel{i\neq j=1}
{l>m}}^N\bar\xi_j\varphi_j({\bf r}_{j})\varphi_{lm}({\bf r}_{lm})\nonumber\\
&=&   \alpha_i k_i {\Psi}
({\bf r}_{1},\cdots,r_i=0,\cdots,{\bf r}_{N}), \ \epsilon_{ilm}\neq 0
.\label{cuspend}\end{eqnarray}
In deriving Eq.\  (\ref{cuspend}) we made use of the fact that
in the limit $(r_i/r_{ij}\to 0)$ the distance $r_{ij}$ tends to $r_j$.
The proof that the wave function (\ref{n3c}) fulfills the cusp conditions
at the collision points of two continuum particles ($r_{ji}\to 0$) runs along
the same lines. Finally, we remark that the wave function (\ref{n3c})  is not
compatible with the expansion of the exact solution of the Schr\"odinger
 equation
(\ref{schra}) at the three-body collision points (e.g.
$ r_i\to 0$ and $r_j\to 0, j\neq i$) since in this case the exact wave function
is known to satisfy a Fock expansion
\cite{fock} in the coordinate $\rho:=\sqrt{(r_i^2+r_j^2)}$
which contains, in addition to powers in $\rho$, logarithmic terms in $\rho$
whereas the wave function (\ref{n3c}) possesses a regular power-series expansion
around $r_i\to 0$ and $r_j\to 0$.
\subsection{Normalization}
The knowledge of the
 normalization factor
 ${\cal N}$ of the wave function (\ref{n3c}) is imperative for the evaluation
of scattering amplitudes
 using the wave function (\ref{n3c}) as a representation
of scattering states.
 In principle, ${\cal N}$ is derived from a $3N$-dimensional
integral over the norm of the function (\ref{n3c}) which, for large $N$,
is an inaccessible task.
Thus for the determination of ${\cal N}$ we resort to the requirement that
the flux through an asymptotic manifold  defined by a constant
 large inter-particle
separations should be the same in the case of
 the wave function (\ref{n3c}) and a
normalized plane-wave representation of the scattering state, i.\ e.\ 
\begin{eqnarray}
{\bf J}_{PW}={\bf J}_{\Psi}
\label{flux}\end{eqnarray}
where the plane-wave flux is given by
\begin{eqnarray}
{\bf J}_{PW}&=&-\frac{i}{2}
 (2\pi)^{-3N}\left[\prod_l^N\bar\xi^*_l({\bf r}_{l}){\bf \nabla}_{}
\prod_l^N\bar\xi_l({\bf r}_{l}) - \prod_l^N\bar\xi_l({\bf r}_{l}){\bf \nabla}_{}
\prod_l^N\bar\xi^*_l({\bf r}_{l})\right]\nonumber\\
&=&(2\pi)^{-3N}\sum_{l=1}^N{\bf k}_{l}\label{pwf}.
\end{eqnarray}
In Eq.\ (\ref{pwf}) the total gradient
${\bf \nabla}_{}:=\sum_{l=1}^N{\bf \nabla}_{l}$ has been introduced.
 To evaluate the flux generated by the
wave function (\ref{n3c}) we note that, by taking advantage of
Eqs.\ (\ref{f12},\ref{phi2nab}),
 we can write for the total gradient of the wave function
(\ref{n3c})
\begin{eqnarray}
{\bf \nabla}_{}\Psi:=P{\cal
 N}\sum_{m=1}^N&&\left\{ i{\bf k}_{m}\Psi
 + \alpha_mk_m{\bf F}_m{\Psi} \ + \ 
        \left[ \sum_{n=m+1}^N
 \alpha_{mn}k_{mn}\overline{\bf F}_{mn}({\bf r}_{mn})
            \prod_{\stackrel{j>i}{j\neq n}}^N\varphi_{ij}
\right.\right.\nonumber\\
& & \left.\left.-\sum_{l=1}^{m-1}\alpha_{lm}k_{lm}
\overline{\bf F}_{lm}({\bf r}_{lm}) 
           \prod_{\stackrel{j>i}{i\neq l}}^N\varphi_{ij}\right]
\prod_{s=1}^N\bar\xi_s({\bf r}_{s})\varphi_s({\bf r}_{s})\right\}
\label{normhlp}\end{eqnarray}
where $\overline{\bf F}_{mn}$ is given by
${\bf F}_{mn}\varphi_{mn}$. The decisive point
now  is that since we are considering
 the flux at large interparticle distances
only the first term of Eq.\  (\ref{normhlp}) is relevant. This is readily deduced from 
Eqs.\ (\ref{bruch1},\ref{bruch12}) which state that all other terms in
Eq.\  (\ref{normhlp}),
except for the first term, can be neglected asymptotically. Note in this context that
terms in the wave function which are 
asymptotically  of the order ${\cal O}(1/r_j,1/r_{lm})$
correspond to parts of
 the Hamiltonian falling off faster than the Coulomb potentials
and hence can be disregarded in the asymptotic regime.
 Now making use of the asymptotic expansion
of the 
confluent hypergeometric function \cite{abra} and taking leading order in the
interparticle distances the flux ${\bf J}_{\Psi}$ can de deduced
\begin{eqnarray}
{\bf J}_{\Psi}={\cal N}^2\prod_{j=1}^N\frac{\exp(\pi\alpha_j)}
{\Gamma(1-i\alpha_j)\Gamma^*(1-i\alpha_j)}\prod_{m>l=1}^N
\frac{\exp(\pi\alpha_{lm})}
{\Gamma(1-i\alpha_{lm})\Gamma^*(1-i\alpha_{lm})}\sum_{n=1}^N{\bf k}_{n}
\label{n2}\end{eqnarray}
where $\Gamma(x)$ is the
 Gamma function. From Eqs.\ (\ref{flux},\ref{pwf},\ref{n2})
it follows that
\begin{eqnarray}
{\cal N}=(2\pi)^{-3N/2}\prod_{j=1,m>l=1}^N\exp[-\pi(\alpha_{lm}+\alpha_{j})/2]
\Gamma(1-i\alpha_{j})\Gamma(1-i\alpha_{lm}).
\label{nend}\end{eqnarray}
For two charged particles moving in the field of a heavy nucleus
the wave function (\ref{n3c}) with
the normalization, given by Eq.\  (\ref{nend}),
simplifies to the three-body wave function
that has been discussed in the previous section.
%
%
%
%
%
%
%
\section{Green function theory of finite correlated systems}
In the preceding sections we investigated the two, three and $N$-body
correlated scattering states. With increasing number of particles the treatment becomes
more complex and a  methodology different from the wave function
technique is more appropriate. A method which is widely used in
theoretical physics is the Green function approach which we will
follow up in this section.

For a   canonical ensemble,
we seek a non-perturbative method which allows to $ $ distribute
systematically the total energy between the potential and the kinetic
energy parts. This is achieved by the development of an incremental
method in which the $N$ correlated particle system is mapped exactly onto a set
of systems in which only $N-M$
 particles are interacting ($M\in [1,N-2]$), i.e. in which
the potential energy part is damped. (In contrast to re-normalization group
theory we do not reduce the strength of interactions, but the number of them).
This is particularly interesting from a thermodynamic point of
view
 since for a number of thermodynamic properties the kinetic energy contributions
can be separated out from the potential energy parts, as shown in the next section
for the internal energy.
By virtue of the present method the potential energy part is systematically reduced.

For a formal development let us consider a
 nonrelativistic system  consisting of $N$ interacting particles.
We assume the total potential  to
be of the  class $U^{(N)}=\sum_{j>i=1}^N v_{ij} $
 without any further specification
of the individual potentials $v_{ij}$. For three-body potentials
the development of the theory proceeds along the same lines.\\

 The  potential $U^{(N)}$ satisfies the recurrence relations
\begin{eqnarray}
U^{(N)}&=&\frac{1}{N-2}\sum_{j=1}^N u_j^{(N-1)},\label{un}\\
u_j^{(N-1)}&=&\frac{1}{N-3}\sum_{k=1}^{N-1}
 u^{(N-2)}_{jk},\,\, j\neq k\label{un1},\end{eqnarray}
 where
$u_j^{(N-1)}$ is the total potential
 of a system of  $N-1$ interacting particles
in which the $j$ particle is missing, i.e.~in terms of the physical
pair potentials $v_{mn}$,
 one can
write
$u_j^{(N-1)}=\sum_{m>n=1}^N v_{mn}, \: m\neq j\neq n $.

The fundamental quantity that describes the microscopic properties
 of the $N$ body quantum system is the
Green operator $G^{(N)}$ which is the
resolvent of the total Hamiltonian. It can be deduced
from the Lippmann Schwinger
equation $G^{(N)} = G_0 + G_0U^{(N)}G^{(N)}$ where
$G_0$ is the Green operator of the non
interacting
$N$ body system.
An equivalent approach to
 determine the dynamical behavior of a system is to  derive the
respective transition operator $T^{(N)}$ which satisfies the integral equation
$T^{(N)}=U^{(N)}+U^{(N)}G_0T^{(N)}$.
These integral equations for $G^{(N)}$ and $T^{(N)}$ provide a natural framework for perturbative
treatments. However,
for $N\ge 3$ the application of
 the  above Lippmann Schwinger equations (and those for the state
vectors)
is hampered by mainly two difficulties: 1.) as shown in Refs. \cite{lippmann,foldy}
the Lippmann Schwin\-ger equations for the state vectors do not have a unique
solution, and 2.)
as shown by  Faddeev \cite{fadd1,fadd2,merk} the kernel of these
integral equations $K= G_0U^{(N)}$ is not a square integrable operator for $N\ge 3$,
i.e.\  the norm  $\| K \| =[ {\rm Tr}(KK^\dagger)]^{1/2}$ is  not square
integrable. The kernel $K$ is also not compact. The reason for this drawback is the
 occurrence of the so-called disconnected diagrams where one of the $N$  particles
is a spectator, i.e. not correlated with the other
$N-1$ particles. For the three-body problem Faddeev \cite{fadd1,fadd2}
suggested alternative integral
equations with square integrable kernel.

Our aim here is twofold: {\bf (a)} We would like to derive non-perturbative
integral equations that treat all $N$ particles on equal footing and are free from disconnected diagrams.
{\bf (b)} These equations should
allow to obtain, in  a computationally accessible manner,
the  solution of the  correlated $N$ body problem
 from the solution when only $N-M$ particles are interacting (where
$M\in [1,N-2]$).

According to the decomposition (\ref{un}), the integral equation for the transition operator can be
written as
\begin{eqnarray}
T^{(N)}&=&\sum_{j=1}^NT^{(N-1)}_j\label{tn}\\
T^{(N-1)}_j&=&\tilde u_j^{(N-1)} +
T^{(N)}G_0\tilde u_j^{(N-1)}, \, \, j\in [1,N].\nonumber\\
\label{tn1} \end{eqnarray}
Here we introduced the scaled potential
\[ \tilde u_j^{(N-1)}=
\left(u_j^{(N-1)}\right)/(N-2). \]

 The transition operator of the system, when
$N-1$   particles are  interacting  via the  scaled potential $\tilde u_j^{(N-1)}$,
is
\[ t^{(N-1)}_j= \tilde u_j^{(N-1)} +  \tilde  u_j^{(N-1)} G_0t^{(N-1)}_j.\]

  With this relation Eq.(\ref{tn1}) can  be reformulated as
\begin{eqnarray}
T^{(N-1)}_j&=&  t^{(N-1)}_j +  t^{(N-1)}_jG_0 T^{(N)} -\nonumber\\
&&
t^{(N-1)}_jG_0\left(\tilde u_j^{(N-1)} +\tilde u_j^{(N-1)} G_0 T^{(N)}\right)\nonumber\\
&=&  t^{(N-1)}_j +  t^{(N-1)}_jG_0 \left( T^{(N)}-T^{(N-1)}_j \right)\nonumber\\
& =&
 t^{(N-1)}_j +  t^{(N-1)}_jG_0\sum_{k\neq j}^N T^{(N-1)}_k.\nonumber\\
\label{ex}
\end{eqnarray}
Eq.(\ref{ex}) can be expressed in a matrix form as follows
\begin{eqnarray}
\left( \begin{array}{c} T^{(N-1)}_1 \\
T^{(N-1)}_2\\
\vdots\\
T^{(N-1)}_{N-1}\\ T^{(N-1)}_{N} \end{array}\right)
&=&\left( \begin{array}{c} t^{(N-1)}_1 \\
t^{(N-1)}_2\\
\vdots\\
t^{(N-1)}_{N-1}\\ t^{(N-1)}_{N} \end{array}\right) 
+
[{\bf K}^{(N-1)}] \left( \begin{array}{c} T^{(N-1)}_1 \\
T^{(N-1)}_2\\
\vdots\\
T^{(N-1)}_{N-1}\\ T^{(N-1)}_{N} \end{array}\right).
\label{expan}\end{eqnarray}
The kernel $[{\bf K}^{(N-1)}]$ is a  matrix operator and is given by
\begin{eqnarray}
[{\bf K}^{(N-1)}]=
\left( \begin{array}{ccccc}
 0          & t^{(N-1)}_1 & t^{(N-1)}_1&\multicolumn{1}{c}\dotfill & t^{(N-1)}_1\\
t^{(N-1)}_2 & 0    & t^{(N-1)}_2& \multicolumn{1}{c}\dotfill   &  t^{(N-1)}_2 \\
\multicolumn{5}{c}\dotfill\\
t^{(N-1)}_{N-1}&    \multicolumn{1}{c}\dotfill  &t^{(N-1)}_{N-1}  &  0 &t^{(N-1)}_{N-1}\\
t^{(N-1)}_{N}  &   \multicolumn{1}{c}\dotfill   & t^{(N-1)}_{N} &   t^{(N-1)}_{N} & 0
 \end{array} \right) G_0. \nonumber\\
\end{eqnarray}
From Eq.(\ref{un1}) it is clear that $t_j^{(N-1)}$
can also be expressed in terms of the transition operators  of the system
where  only $N-2$  particles are interacting:
\[t_j^{(N-1)}=\sum_{k\neq j}^{N-1}T^{(N-2)}_k.\]
 The operators $T^{(N-2)}_k$ are deduced from
Eq.(\ref{expan}) with $N$ being replaced by $N-1$.

From the relation $G^{(N)} = G_0 + G_0 T^{(N)} G_0$ we conclude that
 the Green operator of the interacting $N$ particle system has the form
 \begin{equation}
G^{(N)}=G_0 + \sum_{j=1}^NG^{(N-1)}_j.
\label{greenn}\end{equation}
 The operators
 $G^{(N-1)}_j$ are related to the  Green operators  $g^{(N-1)}_j$ of the systems
in which only $N-1$ particles are correlated  by virtue of $\tilde
u_j^{(N-1)}$. This interrelation is given  via 
\begin{eqnarray}
\left( \begin{array}{c} G^{(N-1)}_1 \\
G^{(N-1)}_2\\
\vdots\\
G^{(N-1)}_{N-1}\\ G^{(N-1)}_{N} \end{array}\right)
&=&\left( \begin{array}{c} g^{(N-1)}_1 -G_0\\
g^{(N-1)}_2-G_0\\
\vdots\\
g^{(N-1)}_{N-1}-G_0\\ g^{(N-1)}_{N} -G_0\end{array}\right)
+
[{\bf \tilde K}^{(N-1)}] \left( \begin{array}{c} G^{(N-1)}_1 \\
G^{(N-1)}_2\\
\vdots\\
G^{(N-1)}_{N-1}\\ G^{(N-1)}_{N} \end{array}\right),
\label{expang}\end{eqnarray}

where $[{\bf \tilde K}^{(N-1)}] = G_0[{\bf \tilde K}^{(N-1)}] G_0^{-1}$.
 From Eqs.(\ref{expan},\ref{expang}) we conclude that if the Green operator of the
interacting $N-1$ body system  is known (from other analytical
or numerical procedures, e.g. from an effective field method, such as density functional
theory)
 the Green operator of the $N$
particles
can then
be deduced by solving a set of $N$ linear, coupled integral equations
(namely
Eqs.(\ref{expan},\ref{expang})).  According to the above equations,
if only the solution of the $N-M$ problem is known
where $M\in[1,N-2]$ we have to perform a hierarchy of calculations
 starting by obtaining the
solution for the $N-M+1$ problem and repeating the procedure to reach the solution of the
$N$ body problem.

At first sight the kernels of Eqs. (\ref{expan},\ref{expang}) appear to have disconnected
diagrams since they contain  transition operators of systems where only
 $N-1$ particles are interacting and   one particle is free (disconnected).
 It is, however, straightforward to show that any iteration of these kernels is free of disconnected
terms (the disconnected terms occurs only in the off-diagonal elements of $[{\bf K}^{N-M}]$ and
$[\tilde {\bf K}^{N-M}]$).
For $N=3$ the present scheme reduces to the  well-established Faddeev equations.
As for the functional structure of the Eqs. (\ref{expan},\ref{expang}) we remark that
 for the solution of the $N$ particle problem we need the (off-shell)
transition operators of the $N-1$ subsystem. The interaction potentials
do not appear in this formulation (in contrast to the Lippmann Schwinger approach).
 On the other hand the (on-shell) transition matrix elements can be determined
experimentally. This fact becomes
 valuable when the potentials are not known.
\subsection{Application to four-body systems}
Over the years a substantial body of knowledge on the three-particle problem has been accumulated.
In contrast,
 theoretical
 studies on the four-body problem are still scare due to computational
limitations whereas an impressive amount of  experimental data is already available
\cite{wehlitz,azzi,alex,robert1,marji}.
Thus, it is  desirable
 to apply the above procedure to the four-body system  and to
express its solution in
terms of known solutions of the three-body problem. For $N=4$
the first iteration of Eq.(\ref{expang})
yields
\begin{equation}
G^{(4)}= \sum_{j=1}^4 g^{(3)}_j -3 G_0
.\label{g4}\end{equation}
Here $g^{(3)}_j$ is the Green operator of the system where only
 three particles are  interacting and can be
taken from other numerical or analytical studies. This means, to a first order, methods
treating  the correlated three-body problem can be extended
to deal with the four-body case using Eq.(\ref{g4}).
We note that for the case of non-interacting system $g^{(3)}_j$ reduces
to $g^{(3)}_j\equiv G_0$ and hence Eq.(\ref{g4}) reduces to $G^{(4)}=G_0$, as
expected.

 The Green function  encompasses the complete spectrum of the
many-body system, i.e.~the wave function approach can be retrieved
from the Green function. For example,  Eq.(\ref{g4}) leads to an
 expression for the four-body state vector in the form
\begin{eqnarray}
|\Psi^{(4)}\rangle &=&|\psi^{(3)}_{234}\rangle +|\psi^{(3)}_
{134}\rangle +|\psi^{(3)}_{124}\rangle +|\psi^{(3)}_{123}\rangle 
- 3 |\phi^{(4)}_{\rm
free}\rangle
.\label{psi}\end{eqnarray}
Here $|\psi^{(3)}_{ijk}\rangle$ is the state vector of the system in which the
   three particles
$i,j$ and $k$ are interacting whereas
$ |\phi^{(4)}_{\rm
free}\rangle$ is the state vector of the non-interacting
four-body system.
The state vectors $|\psi^{(3)}_{ijk}\rangle$ can be approximated
by Eq.(\ref{ds3c}) or by the other procedures discussed in the
preceding section on the three-body problem.

 Since the state vector (\ref{psi}) is expressed as a
 sum of correlated three-body states,
the evaluation of the
 four-body transition matrix elements for a specific
 reaction simplifies considerably.
In addition, the spectral properties of a many-body interacting
system can be obtained
in a straightforward way from those for systems with a reduced number of interactions,
for in this case the matrix elements of the total Green functions
are  expressed as sums of matrix elements of  reduced Green functions,
as evident from Eq.(\ref{g4}). This spectral feature
 can be exploited to study  the thermodynamical
properties of finite correlated systems.
\section{Thermodynamics and phase transitions of interacting finite systems}
To investigate the thermodynamical properties of $N$ interacting particle
system we remark that at the critical point divergent thermodynamical quantities,
such as the specific heat $C_V$ are obtained as a derivative with respect
to the inverse temperature $\beta$ of the logarithm of the
canonical partition function $Z(\beta)$,
\[
C_V=\beta^2\partial ^2_\beta \ln Z(\beta)=f(\beta,Z(\beta))/Z(\beta).\]
Here $f$ is some analytical function and   for the Boltzmann constant we assume $k=1$.

Therefore divergences in the thermodynamic quantities, which
signify  phase transitions are connected to the zero points of
$Z(\beta)$. These zero points are generally complex valued.
Therefore an analytical continuation of $Z(\beta)$ to complex
temperatures is needed.

The connection between the phase transitions and the complex zero points of the
grand canonical partition function have been uncovered by Yang and Lee  \cite{yang}.
 In this case one seeks an analytical continuation of the fugacity
$z=\exp(\beta\mu)$ (here $\mu$ is the chemical potential)
 to the complex plane $z\to \Re(z)+i\Im(z)$. In the
thermodynamic limit the zero points condense to lines. The transition points
are the crossing points of these lines with the
real fugacity axis.

Grossman {\it et al.} \cite{gross} generalized the concept of
Yang and Lee to the canonical ensemble. In this case the inverse
temperature $\beta=1/T$ is continued analytically to
$\beta=\Re(\beta)+i\Im(\beta)$. The phase transitions are then the
crossings of the zero points line of $Z(\beta)$ with the real
$\beta$ axis. The advantage here is that a classification of the
phase transitions can be given in terms of how the zero-points
line do cross the real $\beta$ axis \cite{gross}.

The crucial point  is that in the thermodynamic limit
 $N\to \infty,\, V\to \infty$
and $v=V/N <\infty$ ($V$ is the volume)
the zero points approach, to an infinitesimal small distance
the real axis. For this reason, the characteristic phase-transition
divergences  appear in  the thermodynamical
quantities.
For finite systems $Z(\beta)$ has only finite zero points that
do not approach infinitely close the real axis. Therefore, the thermodynamic
quantities show smooth peaks rather then divergences. The positions and widths
of these peaks can be obtained from the real and imaginary parts of the
zero points laying closest to the real axis \cite{barber}.

To apply this method to correlated finite systems we need a
representation of the canonical partition function that can then
be continued analytically to the complex temperature plane.

The canonical partition function of a correlated system
can be expressed in terms of the many-body Green function as
\begin{equation}
Z(\beta)=\int d E\; \Omega(E)\; e^{-\beta E}
.\label{eqq1}
\end{equation}
Here $\Omega(E)$ is the density of states which is  related to
 imaginary part of the trace of  $G^{(N)}$ via
\begin{equation}
 \Omega(E)=-\frac{1}{\pi}\Im\, {\mathbf{ Tr}}\, G^{(N)}(E).
\label{eqq2}\end{equation}
From the Green function expansion Eq.(\ref{greenn})
we deduce
\begin{eqnarray}
Z(\beta)&=&-\frac{1}{\pi}\Im\int d E\;
 {\mathbf{ Tr}}\, G^{(N)}(E)\; e^{-\beta E}\nonumber\\
&=& Z_0(\beta)+\sum_{j=1}^N Z_j(\beta)
\label{eqq21}
\end{eqnarray}
where
\begin{eqnarray}
Z_0(\beta)&=&-\frac{1}{\pi}\Im\int d E\;
 {\mathbf{ Tr}}\,  G_0(E)\; e^{-\beta E}\\
Z_j(\beta)&=&-\frac{1}{\pi}\Im\int d E\;
 {\mathbf{ Tr}}\, G_j^{(N-1)}(E)\; e^{-\beta E}
.\nonumber\\
\label{eqq3}
\end{eqnarray}
To a first order $Z_j=Z_j^{(N-1)} -Z_0$ where $Z_j^{(N-1)}$
is the partition function of a system in which only $N-1$ particles
are interacting.
For the  applications of the Grossmann method let us remark
that  $Z(\beta)$ is an integral function and
can be expressed in a polynomial form. Recalling the analytical properties of
meromorphic functions one can write $Z(\beta)$ in terms
of its complex zero points
as
\begin{equation}
Z(\beta)=Z(0) e^{\beta\frac{Z'(0)}{Z(0)}}\: \prod^{\infty}_{k=1} \left(1-\frac{\beta}{\beta_k}
\right) e^{\frac{\beta}{\beta_k}}.
\label{eqq4}\end{equation}
\section{Conclusions}
 In conclusions we reviewed recent advances  in the  analytical  treatment
of correlated few charged-particle  systems.
Starting from the two-body (Kepler) problem we discussed
the mathematical and physical properties of  a three-body system
above the complete fragmentation threshold.
Analytical approximate expressions for the three-body wave functions
have been derived and employed for the calculations of transition
matrix elements in atomic scattering experiments.
The discussion has been extended to the case of $N$ charged particles
in the continuum where $N$ particle wave functions have been derived
and their mathematical features have been exposed. 
For the description of the complete spectrum of a general finite system we discussed a
Green function method that maps the  $N$ interacting particle system onto a
system with a less number of interactions. A brief account has been given on how this method
can be used to investigate thermodynamic properties of finite systems.
The application  of the Green function method for the calculations of scattering
amplitudes  have been presented in Ref.\cite{prlfadd}.  The
 Green function method presented here have been  extended recently to deal with
the scattering of correlated systems from ordered and disordered potentials \cite{srl00}.

\end{document}